\documentclass[sigconf]{acmart} 

\copyrightyear{2026}
\acmYear{2026}
\setcopyright{cc}
\setcctype{by}
\acmConference[ICSE '26]{2026 IEEE/ACM 48th International Conference on Software Engineering}{April 12--18, 2026}{Rio de Janeiro, Brazil}
\acmBooktitle{2026 IEEE/ACM 48th International Conference on Software Engineering (ICSE '26), April 12--18, 2026, Rio de Janeiro, Brazil}
\acmPrice{}
\acmDOI{10.1145/3744916.3773175}
\acmISBN{979-8-4007-2025-3/26/04}

\usepackage{multirow}
\usepackage{listings}

\usepackage[ruled,vlined,linesnumbered,lined]{algorithm2e}
\usepackage[shortlabels]{enumitem}

\usepackage{wrapfig}
\usepackage[table]{xcolor}% http://ctan.org/pkg/xcolor
\usepackage{algpseudocode}
\usepackage{comment}

\newsavebox{\answerboxbox}
\newenvironment{answerbox}{%
  \begin{lrbox}{\answerboxbox}%
    \begin{minipage}{0.97\linewidth}%
    \ignorespaces
}{%
    \end{minipage}%
  \end{lrbox}%
  \noindent\fcolorbox{gray!50}{gray!5}{\usebox{\answerboxbox}}%
}

\newsavebox{\grayframebox}

\author{Verya Monjezi}
\orcid{0000-0001-6796-5948}
\affiliation{%
  \institution{University of Illinois Chicago}
  \city{Chicago, IL}
  \country{USA}
}
\email{vmonj@uic.edu}

\author{Ashish Kumar}
\orcid{0000-0001-8773-2084}
\affiliation{%
  \institution{Penn State}
  \city{State College, PA}
  \country{USA}
}
\email{azk640@psu.edu}

\author{Ashutosh Trivedi}
\orcid{0000-0001-9346-0126}
\affiliation{%
  \institution{University of Colorado Boulder}
  \city{Boulder, CO}
  \country{USA}
}
\email{ashutosh.trivedi@colorado.edu}

\author{Gang Tan}
\orcid{0000-0001-6109-6091}
\affiliation{%
  \institution{Penn State}
  \city{State College, PA}
  \country{USA}
}
\email{gtan@psu.edu}

\author{Saeid Tizpaz-Niari}
\orcid{0000-0002-1375-3154}
\affiliation{%
  \institution{University of Illinois Chicago}
  \city{Chicago, IL}
  \country{USA}
}
\email{saeid@uic.edu}

\begin{document}
% https://chatgpt.com/share/67cb7725-3cec-8007-ac61-647ee3c8508e
% https://figshare.com/s/2a8f2f92da9858973633

% \title{\scalebox{0.9}{Are Fairness \replaced{Practices}{Design Patterns}  Robust? A Causal \deleted{Debugging} Framework for Testing Robustness}}
\title{On the Robustness of Fairness Practices: \\ A Causal Framework for Systematic Evaluation}

% \author{
% \IEEEauthorblockN{Verya Monjezi}
% \IEEEauthorblockA{
% vmonjezi@miners.utep.edu \\
% \textit{University of Texas at El Paso}}

% \IEEEtitleabstractindextext{
\begin{abstract}
Machine learning (ML) algorithms are increasingly deployed to make critical decisions in socioeconomic applications such as finance, criminal justice, and autonomous driving. 
However, due to their data-driven and pattern-seeking nature, ML algorithms may develop decision logic that disproportionately distributes opportunities, benefits, resources, or information among different population groups, potentially harming marginalized communities.
In response to such fairness concerns, the software engineering and ML communities have made significant efforts to establish the best practices for creating fair ML software. These include fairness interventions for training ML models, such as including sensitive features, selecting non-sensitive attributes, and applying bias mitigators.
But how reliably can software professionals tasked with developing data-driven systems depend on these recommendations? And how well do these practices generalize in the presence of faulty labels, missing data, or distribution shifts? These questions form the core theme of this paper.

We present a testing tool and technique based on causality theory to assess the robustness of best practices in fair ML software development.
Given a practice---specified as a first-order logic property--- and a socio-critical dataset that satisfies the property, our goal is to search for neighborhood datasets to determine whether the property continues to hold. This process is akin to testing the robustness of a neural network for image classification, except that the ``image" is an entire dataset, and its ``neighbors" are datasets in which certain causal hypotheses are altered.
Since computing neighborhood datasets while accounting for various factors---such as noise, faulty labeling, and demographic shifts---is challenging, we utilize causal graph representations of the dataset and leverage a search algorithm to explore equivalent causal graphs to generate datasets.
Our results across various fairness-sensitive tasks, derived from prevalent fairness-sensitive applications, identify best practices that preserve robustness under the varying factors.
\end{abstract}




\maketitle

\section{Introduction}
\label{sec:intro}
Software professionals are increasingly tasked with developing data-driven software systems with socioeconomic and legal implications.
Unlike classical software analysis, detecting fairness vulnerabilities in such systems requires expertise that extends beyond technical competence and domain knowledge. Understanding fairness and discriminatory bugs necessitates a nuanced grasp of demographics, societal structures, systemic biases, social policy, and law.
As a result, the software and ML engineering communities have made concentrated efforts to refine their understanding by proposing various software fairness characterizations and tools. These encompass a wide range of use cases, from individual and group fairness to quantitative and counterfactual fairness.
Recently, there has been a growing trend toward establishing best fairness practices~\cite{zhang2021ignorance,10.1145/3468264.3468536,chakraborty2020fairway,chakraborty2019software,10.1145/3510003.3510202} for ML software to facilitate the transfer of insights from one setting to another. This paper aims to develop a systematic approach to evaluate the robustness of these guidelines.

\vspace{0.25 em}
\noindent\textbf{ Fairness Practices.}
% \vmsays{More practices.}\stsays{The current ones are good enough.}
\textit{Pre-processing.} Zhang and Harman, in their ICSE'21 finding~\cite{zhang2021ignorance}, made a critical observation about the pre-processing during training that \underline{enlarging the feature space} of the dataset during training \emph{can improve fairness}, while increasing the samples size \emph{does not affect fairness} of software.
Biswas and Rajan~\cite{10.1145/3468264.3468536} discuss strategies on data pre-processing (e.g., different data standardization, feature selection, and over/under-sampling operators) and confirm that \underline{selecting a subset of features} often increases unfairness, but the effect depends on the type of operator.
\textit{In-processing.} \textsc{Fairway}~\cite{chakraborty2020fairway,chakraborty2019software} emphasized the role of \underline{hyperparameter} \underline{tuning} in \emph{mitigating the bias}. 
Tizpaz-Niari et al.~\cite{10.1145/3510003.3510202} discovered that certain \underline{hyperparameter configurations} (e.g., \texttt{max\_feature} hyperparameter in decision tree
and random forest classifiers) can consistently introduce \emph{fairness bugs} in the data-driven software. 
\textit{Post-processing.} Hardt et al.~\cite{hardt2016equality} proposed using different \underline{decision} \underline{thresholds} for different groups, and Pleiss et al.~\cite{pleiss2017fairness} \underline{calibrated} \underline{favorable} \underline{outcomes} while minimizing error disparity across different population groups. 

% \vspace{0.25 em}
% \noindent\textbf{Research Questions.}
% Do these design patterns remain locally robust and generalize? How should software developers in fairness-sensitive setting interpret and use these patterns? 
% Besides, how do these patterns remain robust from one social and cultural context to another under a distribution shift?

% ~\vmsays{Only talked about neighbor datasets, not generative models}
% \stsays{the current text looks good to me.}
\vspace{0.25 em}
\noindent\textbf{Research Challenge and Main Idea.}
We posit that robust fairness practices should yield consistent outcomes when applied to neighboring datasets---datasets that are similar but not identical. 
A normative example of such neighboring datasets is the case of gender bias in graduate admissions~\cite{bickel1975sex}, where researchers debate whether sex is an influencing factor in graduate program admissions, indicating systemic bias in the admission process, or whether the choice of program is influenced by the candidate's sex, suggesting a social-level bias.
In either interpretation, fairness practices in ML should remain valid and useful despite variations in data contexts or the underlying relationships between variables such as sex and admissions.
% We hypothesize that if these fairness practices are robust, they should yield consistent outcomes when applied to neighboring datasets. A normative
% example is the sex bias in the grad admissions~\cite{bickel1975sex} where researchers dispute whether sex is a trait of the grad program admission, or the program is a trait of sex. If sex influences admission decisions, 
% it points to systemic bias and discrimination in the admission process. Conversely, if the choice of program is influenced by the candidate's sex, 
% reflecting social norms, it suggests a social-level bias indirectly affecting admissions.\stsays{I think what we want to say is that under both interpretations of relationships between sex and admissions as well as the distribution shifts, explained next, the best practices in the fair ML software should be still valid and useful.}
% \stsays{if there is a time, let us take a look at the works by Moritz Hardt who discuss this as epistemological problem in the causal modeling for fairness. But, idk which work! If we find out, we can position it better.}\vmsays{Causal Inference out of Control: Estimating the Steerability of Consumption}
These scenarios highlight the necessity for fairness practices to be ``robust" to different interpretations and distribution shifts.
Our goal is to examine the ``robustness" of best-practice guidelines in both in-distribution and out-of-distribution scenarios.
Focusing on robustness is crucial for ensuring generalization in real-world applications and establishing fairness best practices.
% These robustness requirements are essential for repeatability and replicability, considering that samples represent limited observations over the underlying data distributions and are susceptible to errors in
% sampling, faulty labeling, and imbalanced representations. \stsays{this paragraph is not super clear, especially it is too early to mention two or more generative models. Also, it is not clear how
% admission and distribution shift examples relate to the robustness.}\vmsays{I agree with you. I think the whole paragraph should be revised}

% Drawing inspiration from characterizing the robustness of decisions in supervised learning~\cite{szegedy2013intriguing,papernot2017practical,wang2021adversarial,HKWW17}, it would be highly beneficial to measure the robustness of various fairness-related empirical claims by examining small perturbations in the underlying data distribution.
% A robust claim should not only hold true in a specific training dataset developed under a particular societal context, but it should also remain valid when the dataset is used in a slightly different context with varying relationships between features.

\vspace{0.25 em}
\noindent\textbf{Characterizing Robustness.}
In ML, robustness refers to a model's ability to maintain performance when confronted with uncertainties or adversarial perturbations~\cite{goodfellow2014explaining,carlini2017towards}. A well-known example of robustness research is the discovery that ML classifiers can produce entirely different classifications when exposed to small, human-imperceptible perturbations~\cite{DBLP:journals/corr/SzegedyZSBEGF13}. For instance, a stop sign with imperceptible noise added could be misclassified as a speed limit sign for 45 mph, posing serious safety risks in autonomous driving systems. Such vulnerabilities highlight the critical need for robustness in ML applications for high-stakes domains.

Our work differs from classical robust ML scenarios in two key ways.
First, the category of robustness we consider is broader: rather than focusing on perturbations to individual inputs, we assess robustness against changes to entire datasets.
Second, defining dataset perturbations requires careful consideration. This is particularly important because ML algorithms are designed for generalizability (e.g., through standard training and testing splits). As a result, a naïve definition based on superficial dataset similarities—such as simple perturbations like Gaussian noise—fails to rigorously assess the robustness of fairness practices.
However, this presents a significant challenge, as the underlying generative models are typically unavailable. To address this, we abstract the core structure of the data-generating process by inferring a weighted causal model from the dataset~\cite{carpenter2017stan}.
This approach systematically analyzes and modifies the data generation process---something that is not feasible with generative AI methods such as GANs and VAEs~\cite{pmlr-v157-zhao21a,10.1007/978-3-031-35891-3_26,Zhang2015LearningCF,GANS-synthetic-danial}.

We propose a search-based approach to scale up the discovery of equivalent causal graphs of data with varying fairness implications across different practices.
Specifically, given a partial causal graph inferred by a causal discovery algorithm~\cite{spirtes2000causation,chickering2002optimal,10.5555/3020419.3020473} that contains one or more unresolved (bi-directional) edges, we explore the equivalence classes of graphs. We then introduce perturbations to assess fairness outcomes under different conditions, identifying edge-case scenarios where established fairness practices fail.
To achieve this, we examine various causal graph-theoretic notions of proximity in our search for counterexamples of robustness, allowing us to identify two equivalent causal graphs (with all edges resolved) that yield differing fairness outcomes.
We hypothesize that robustness analysis can uncover subtle perturbations that may not be detectable by analyzing individual datasets alone. This approach provides a more nuanced understanding of the generalizability of fairness findings.
By focusing on neighboring generative models, we gain deeper insights into the robustness of fairness practices and their applicability across diverse contexts. This view is a significant shift from the prevalent fairness testing techniques in the SE literature~\cite{angell2018themis,10.1145/3338906.3338937,agarwal2018automated,udeshi2018automated,10.1145/3510003.3510137,zhang2020white,10.1145/3460319.3464820,9793943}. While the prior work tested ML models for a given fairness metric, our approach tests the robustness of common fairness practices that are broadly applicable for engineering data-driven software beyond a specific model and task.

% \vspace{0.25 em}
% \noindent
% \textbf{Relevance to SE Debugging.} Our approach draws inspiration from delta debugging~\cite{zeller2002simplifying} by systematically testing minimal perturbations to isolate failure conditions. In the context of fairness evaluation, instead of program inputs, we explore neighboring datasets generated via controlled perturbations of causal relationships. Just as delta debugging aims to uncover the minimal subset of input responsible for a program crash, our framework seeks to identify minimal shifts in the data distribution or causal structure that cause a fairness intervention to fail. This leads to a practical debugging workflow: the software engineer already suspects a fairness bug and is evaluating a candidate fairness intervention to mitigate the issue. In this sense, discovering whether a fairness recommendation resolves the issue or introduces new ones can reasonably be seen as a debugging process.

\vspace{0.25 em}
\noindent\textbf{Research Questions.}
In this paper, we aim to experimentally address the following research questions (RQs):

\begin{enumerate}[start=1,label={\bfseries RQ\arabic*},leftmargin=3em]
\item {\bf What is the quality of data generation by different causal discovery algorithms?}
We first use the equivalence causal graphs for these datasets and study the efficacy of various causal discovery algorithms. Our results show that GES~\cite{chickering2002optimal}  outperforms  PC~\cite{spirtes2000causation}, SIMY~\cite{10.5555/3020419.3020473}, and random baseline in generating adversarial neighbor datasets. 

\item {\bf Are the best fairness practices robust when non-sensitive or sensitive attributes are dropped during training with neighborhood causal graphs?} 
We leverage the causal graphs to generate neighborhood datasets and study the robustness of dropping sensitive attributes where we find that it may hold in one dataset but not in the other neighbor dataset. \textit{When analyzing non-sensitive attributes, we observe that \texttt{SelectFpr}~\cite{SelectFpr} (selecting top features based on the false positive rates) demonstrates considerable robustness for both in-distribution and out-of-distribution scenarios.}

\item {\bf Do hyperparameter configurations remain robust w.r.t fairness of outcomes when the underlying causal representations slightly change?}
We perform the same analysis but with different hyperparameters (HPs), as compared to the default, to understand if any configuration may systematically change fairness. 
\textit{Our analysis finds that hyperparameters of logistic regression (LR) classifiers remain robust w.r.t. group fairness when causal graphs are slightly perturbed.}

\item {\bf Are the post-processing bias mitigation practices robust w.r.t fairness? }
We consider two well-established post-processing bias mitigators. We test the robustness of Threshold Optimizer~\cite{hardt2016equality} and Calibrated Equalized Odds~\cite{NIPS2017_b8b9c74a}. Our analysis shows that these practices are not robust in most cases. \textit{We find and report cases where one method remains robust for a dataset across varying training algorithms.}
\end{enumerate}

\vspace{0.5em}
\noindent\textbf{Contributions.}
We observe that the task of studying the robustness of fairness practices is significant because small, controlled variations in the dataset can affect fairness outcomes, and developers may not always be equipped to find and evaluate subtle changes properly. By systematically finding these variations, we aim to identify edge-case situations where the best fairness guidelines may fail. Causal graphs offer a structured way to do this, and our results corroborate that omitting causal graphs underestimates fairness violations. The key contributions of this paper are:
\begin{itemize}[leftmargin=*]
    \item We present a systematic search algorithm on the basis of causality to verify the robustness of various fairness practices;
    % \item We study the performance of three widely used causal graph discovery algorithms in generating realistic data viz-a-viz random graphs and datasets;
    \item We present an automated tool that takes a practice in the pre-processing, in-processing, and post-processing stages as input and quantifies their local robustness; and
    \item We conduct a series of experiments to validate the robustness of eight fairness practices over
    six fairness-sensitive tasks, three training algorithms, and three causal algorithms.
    % \vmsays{should we report the total number of experiments as well?}. \stsays{we do not have to, but if we know the number, it would be nice.}
\end{itemize}

\section{Preliminaries}
\label{sec:background}

\noindent \textbf{Fairness Terminology.}
\label{subsec:Fairness-Terminology}
We consider a data-driven software system with \textit{binary} outcomes where a prediction label is \textit{favorable}
if it outputs a desirable outcome for the target individual.
Examples of favorable predictions are low risks of accidents in insurance applications,
high first-year GPAs in graduate school, and low risks of re-offending in parole assessments.
Each dataset consists of a number of \textit{attributes} (such as income,
experience, prior arrests, sex, and race) and a set
of \textit{instances} that describe the value of attributes for each individual.
According to ethical and legal requirements, data-driven software should not \textit{discriminate}
on the basis of an individual's \emph{protected attributes} such as sex, race, age, disability,
color, creed, national origin, religion, genetic information, marital status, and sexual orientation.

\vspace{0.25 em}
\noindent \textbf{Fairness Metric.}
% There are two types of fairness notions: \textit{individual fairness} and \textit{group fairness}. 
% \textit{Individual fairness}~\cite{dwork2012fairness} often requires that two \textit{individuals} that are deemed similar (based on their non-protected attributes) are treated similarly. 
% Group fairness requires that the statistics of ML outcomes for different \emph{protected groups} to be similar~\cite{hardt2016equality} using metrics such as \textit{equal opportunity difference} (EOD), which is the difference between
% the true positive rates (TPR) of two protected groups, or \textit{average odd difference} (AOD),
% which is the average of differences
% between the true positive rates (TPR) and the false positive rates (FPR) of
% two protected groups~\cite{bellamy2019ai,chakraborty2020fairway,zhang2021ignorance}.
% \textit{Our approach is geared toward group fairness.}
Fairness notions include both \textit{individual} and \textit{group} perspectives. \textit{Individual fairness}~\cite{dwork2012fairness} emphasizes similar treatment for similar \textit{individuals} based on non-protected attributes. Group fairness focuses on achieving similar outcome statistics across different  \emph{protected groups}. Metrics like \textit{equal opportunity difference} (EOD) and \textit{average odds difference} (AOD) quantify disparities in true positive and false positive rates between groups~\cite{bellamy2019ai,chakraborty2020fairway,zhang2021ignorance}. \textit{Our approach is geared toward group fairness since the practices from the literature have used group metrics~\cite{chakraborty2020fairway,chakraborty2019software,10.1145/3510003.3510202,10.1145/3468264.3468536}.}

\vspace{0.25 em}
\noindent \textbf{Designing of Training Process.}
% While conventional software developers directly implement decision-making logic through control and data structures, 
% those working with machine learning (ML) supply the system with input data, perform some preliminary data processing,
% select appropriate ML algorithms, adjust hyperparameters, and perform training. 
% The \textit{training process} allows ML systems to derive a model that encapsulates the decision-making logic
% based on the provided data. As a part of the training process, ML users evaluate the resultant model over
% some \textit{validation set} to measure functional metrics such as overall accuracy (the ratio of correct predictions)
% and F1 score (the mean of precision and recall) as well as non-functional metrics such as EOD for group fairness.
Unlike traditional software development, ML systems derive decision-making logic through a \textit{training process}. This involves providing input data, selecting algorithms, adjusting hyperparameters, and iteratively refining a model. Evaluation on a \textit{validation set} assesses functional metrics like accuracy and F1 score, alongside fairness metrics such as EOD.

% Previous works suggest multiple recommendations for fair configurations of training process based on empirical
% observations~\cite{zhang2021ignorance,10.1145/3468264.3468536,dwork2012fairness,10.1145/3510003.3510202}.
% Fairness through awareness~\cite{dwork2012fairness} requires including sensitive attributes during training for a fair data-driven software development. Zhang and Harman~\cite{zhang2021ignorance} reported that limiting the feature space of dataset during training can degrade fairness.
% Biswas and Rajan~\cite{10.1145/3468264.3468536} studied various pre-processing operations in the training process
% and found that some feature selections operators like \texttt{SelectKBest}~\cite{KBest} and \texttt{SelectPercentile}~\cite{SelectPercentile}
% increase biases, while others like \texttt{SelectFpr}~\cite{SelectFpr} do not impact fairness. 
% \textsc{Parfait-ML}~\cite{10.1145/3510003.3510202}
% found that setting some hyperparameters like \texttt{max\_feature} in random forests to specific values can systematically increase biases in the data-driven software.

\vspace{0.25 em}
\noindent \textbf{Causality Analysis.} Causation, or a causal relationship, involves a link between two
variables where alterations in one variable directly affect changes in the other. 
This principle is distinct from correlation, which only signifies a statistical association between two variables,
without implying a direct cause-and-effect relationship. 
Two variables (say the treatment $X$ and the outcome $Y$) can statistically correlate with each other,
but only one of the following cause-effect scenarios holds~\cite{pearl2018book}:
(1) $X$ causes $Y$ (i.e., $X \to Y$); (2) $Y$ causes $X$ ($Y \to X$); and (3) there is a \textit{confounder}
variable $Z$ that causes both $X$ and $Y$ ($X \leftarrow Z \to Y$). Note that associations
between two variables alone cannot distinguish between these scenarios. 
To handle complex cause-effect relationships, we define the causal graph of input variables:
% \begin{definition}[Causal Graph]
  A causal graph is a directed acyclic graph (DAG), made up of vertices $\mathcal{F}$ and edges $E$, denoted as $G$ = ($\mathcal{F}$, $E$). 
  In this representation, each vertex (attribute) $X_1 \in \mathcal{F}$ stands for a random variable, and each edge $X_1 \to X_2$ (for $X_1 \in \mathcal{F}$ and $X_2 \in \mathcal{F}$)
  symbolizes a direct causal link from $X_1$ to $X_2$. There are two types of vertices within the graph:
  endogenous vertices $X \subset \mathcal{F}$, whose values are influenced by other vertices in the graph, and exogenous vertices $\{{U_{Y}}\}$, for $Y \cap X = \emptyset$, whose values are independently generated from some distributions and not influenced by other vertices in the graph.
% \end{definition}
Since causal graphs are not often available, one can use standard causal discovery algorithms to infer the direction
of cause-effect relationships between variables and Bayesian inference algorithms to learn the strengths (weights)
of relationships. 
% Given a causal graph, there are two
% important types of nodes~\cite{pearl2016causal}: confounders and colliders.

% \begin{itemize}
    % \item A configuration with three variables and two edges with one edge directed into and
    % one edge directed out of the middle variable (e.g., $X_1 \to X_2 \to Y$) is called a \textit{chain}. In this case,
    % $X_1$ and $Y$ are conditionally independent given $X_2$ (e.g., when including $X_2$ in training). The middle
    % variable is called a \textit{mediator}. 
    
%     \item A configuration with three variables and two edges with two arrows emanating from the middle
%     variable (e.g., $X_1 \leftarrow X_2 \to Y$) is called a fork. The path between $X_1$ and $Y$ 
%     are blocked by $X_2$ (i.e., given $X_2$, $X_1$ and $Y$ are conditionally independent). The middle
%     variable is called a \textit{confounder}.

%     \item A configuration with three variables and two edges with one node receives edges from two other
%     nodes (e.g., $X_1 \to X_2 \leftarrow Y$) is called a \textit{collider}. In this case, $X_1$ and $Y$ 
%     are independent, but conditioning on $X_2$ opens a \textit{backdoor} path between $X_1$ and $Y$
%     (e.g., given $X_2$, $X_1$ and $Y$ are likely dependent). The middle variable is called a \textit{collider}.
    
% \end{itemize}

\section{Overview}
\label{sec:overview}

\begin{figure}
    \centering
    \includegraphics[width=.49\textwidth]{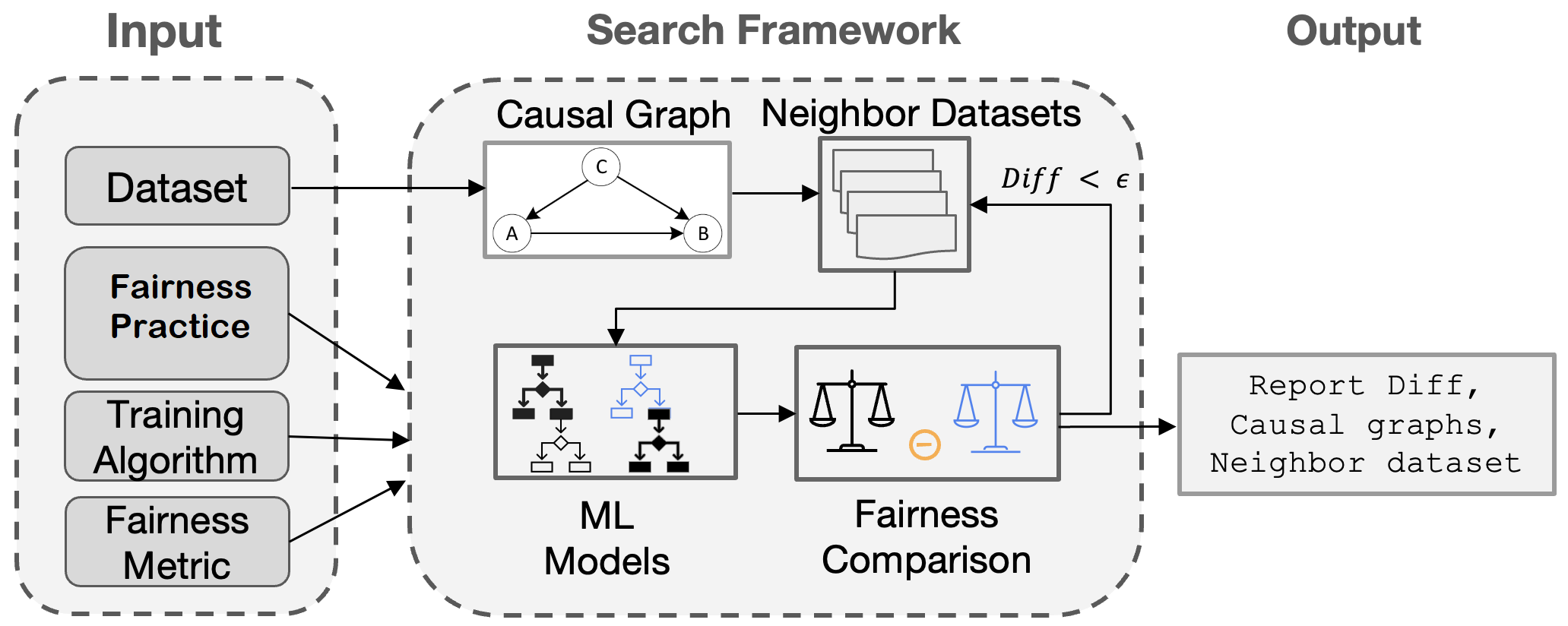}
    \caption{Causal Framework for Robust Fairness.}
    \label{fig:overview}
\end{figure}

\noindent \textbf{Framework.}
Figure~\ref{fig:overview} presents an overview of our proposed framework. The framework takes a dataset, a fairness practice, an ML algorithm, and a fairness metric as inputs and decides whether the practice is locally robust w.r.t. the dataset, the algorithm, and the metric. The search mechanism converts the input dataset into a causal graph representation. Using probabilistic programming techniques, it estimates the posterior distributions of (partial) edges in the graph and generates two neighboring datasets in each step. Then, the training algorithm infers two ML models and measures their fairness differences (diff). If the differences exceed a threshold ($\epsilon$), we deem this a violation of robustness and return the causal graphs. Otherwise, we carefully perturb the most promising causal graph and continue the search until we find a violation or a timeout. A robust fairness practice is expected to yield consistent results.

\begin{figure}[t!]
    \centering
    % \subfloat[CPDAG Graph for Adult.]{\includegraphics[width=0.26\textwidth]{Figures/Adult_SIMY_CPDAG.png}} 
    % % \hspace{0.5 em}
    % % \subfloat[Equivalence graph B for Adult.]{\includegraphics[width=0.26\textwidth]{Figures/Adult_SIMY_base_DAG.png}} 
    % % \hspace{0.5 em}
    % % \subfloat[Equivalence graph C for Adult.]{\includegraphics[width=0.26\textwidth]{Figures/Adult_SIMY_pert_DAG.png}}
    % \caption{Causal graphs generated by GES algorithm. Occupation, hours-per-week, and label are shown with o, h, and y. }
    % \label{fig:Adult_DAGs}
    % % \vspace{-1.0 em}

    % \subfloat[Causal Graph for Adult \stsays{remove this}.]
    {\includegraphics[width=0.3\textwidth]{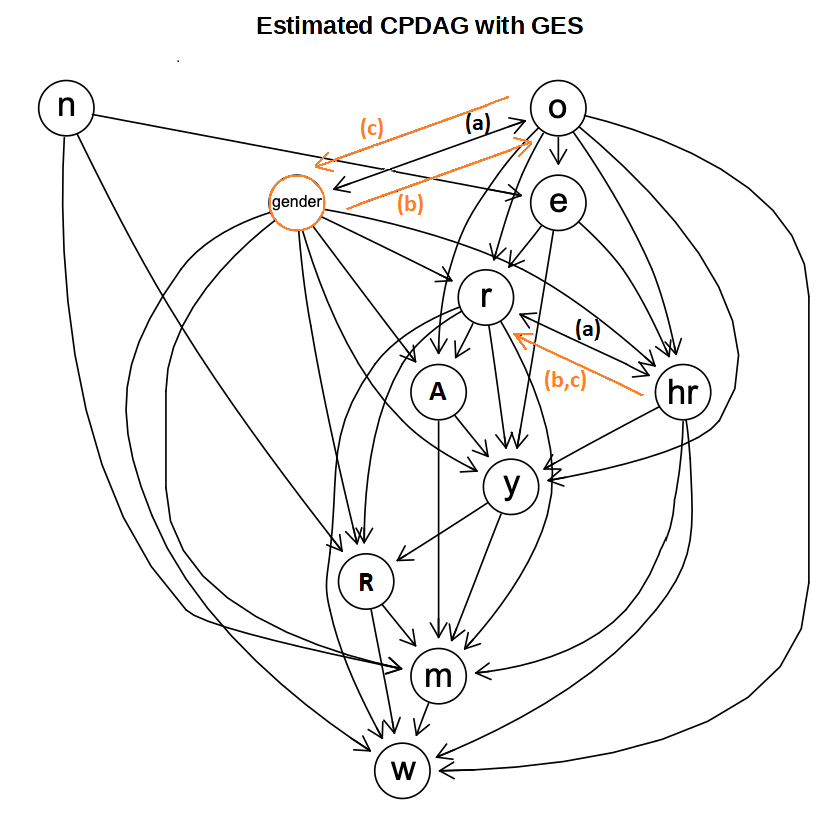}}
    % \subfloat[Fairness Diff in Practice 1 between causal graphs (b) and (c).] 
    % {\includegraphics[width=0.3\textwidth]{Figures/Adult_sens_bopx_plot.png}}

    \captionsetup{font=small}
    \caption{Causal graph (a) generated by GES algorithm with unresolved edges. Causal graphs (b) and (c) are two equivalent DAGs. 
    % \textbf{Right.} Fairness differences of excluding and including sensitive attributes in causal graph (b) shown in top vs. causal graph (c) shown in bottom.
    }
    \label{fig:Adult_DAGs}
    % \label{fig:Adult_sens_drop}
    % \vspace{-1.0 em}
\end{figure}

Next, we overview the robustness of guidelines in developing fair ML software using an example of the Adult Census dataset. 

% \vspace{0.25 em}
% \subsection{Incorporating Causal Graph.} 
\noindent \textbf{Incorporating Causal Graph.} 
The Adult Census dataset consists of ten features. The primary goal of this dataset is to predict whether an individual's income exceeds 50k per year based on personal and demographic details such as \textit{gender}, \textit{race}, \textit{relationship}, and \textit{education level}. We consider \textit{gender} as the sensitive attribute for this task.

We utilized three widely recognized causal discovery algorithms— PC~\cite{spirtes2000causation}, GES~\cite{chickering2002optimal}, and SIMY~\cite{10.5555/3020419.3020473}---to infer the causal structures in the Adult Census dataset. In addressing the challenging problem of causal graph inference from the dataset, we encountered a fundamental challenge inherent in existing causal discovery algorithms, including PC, GES, and SIMY. While effective in identifying possible causal connections between features, these methods frequently fail to produce a single, definitive Directed Acyclic Graph (DAG). Instead, they produce a Completed Partially Directed Acyclic Graph (CPDAG), which is a collection of equivalent DAGs with unresolved directional ambiguity in causal relationships. 

% Figure~\ref{fig:Adult_DAGs} (a) shows the produced CPDAG by the GES algorithm for the Adult dataset. In the figure, we use letters to represent some features due to the readability of the graphs, e.g., the letters o, r, and hr represent occupation, relationship, and hours-per-week, respectively.

% As the figure suggests, the CPDAG contains two bi-directional edges (e.g., gender $\leftrightarrow$ occupation). Each of these bi-directional edges can take either direction, hence the equivalence class contains up to four unique DAGs (highlighted by red arrows), with an assumption of not having an unobserved confounding variable. For instance, Figures~\ref{fig:Adult_DAGs} (b,c) illustrate two different DAGs generated from CPDAG shown in Figure~\ref{fig:Adult_DAGs} (a) where they have the same skeleton. The only difference between (b) and (c) is the edge between gender and occupation (o) where we have gender $\rightarrow$ occupation (o) and gender $\leftarrow$ occupation (o) in Figure~\ref{fig:Adult_DAGs} (b) and (c), respectively. Intuitively, both directions can be valid,
% depending on the ontological interpretations of gender and occupation. With the arrow from gender to occupation, we treat gender as a trait for occupations whereas with the arrow from occupation to gender, we consider occupation as a trait for gender (analogous to the famous sex bias case in the UC Berkeley grad admission~\cite{bickel1975sex}).

Figure~\ref{fig:Adult_DAGs} shows the produced CPDAG by the GES algorithm for the Adult dataset. In the figure, we use letters to represent some features due to the readability of the graphs, e.g., the letters o, r, and hr represent occupation, relationship, and hours-per-week, respectively. The CPDAG contains two bi-directional edges that lead to multiple possible DAG interpretations, hence the equivalence class contains up to four unique DAGs. The highlighted labeled arrows within Figure~\ref{fig:Adult_DAGs} (a) show bi-directional. 
%\vmsays{Does it explain arrows well?} 
For instance, the gender $\leftrightarrow$ occupation edge could represent either gender influencing occupational choices (Figure~\ref{fig:Adult_DAGs} (b)) or occupation shaping societal perceptions of gender (Figure~\ref{fig:Adult_DAGs} (c)). Intuitively, both directions can be valid, depending on the ontological interpretations of gender and occupation. With the arrow from gender to occupation, we treat gender as a trait for occupations whereas with the arrow from occupation to gender, we consider occupation as a trait for gender (analogous to the famous sex bias case in the UC Berkeley grad admission~\cite{bickel1975sex}). This highlights the dynamic relationship between these variables, influenced by societal norms, cultural perceptions, and historical contexts. These factors can evolve, but the best practices in data-driven software should remain effective. 

Given a DAG, we use the Bayesian inference with STAN to infer posterior distributions over the feature and the coefficients of linear models that connect different features.
When generating in-distribution datasets from the causal graphs, we also include a validation step
where we use a clustering of original datasets with a distance function to reject any samples
that are far from any modality in the dataset.
Thus, it ensures that our generated samples remain
representative and within the parameters of realistic data distributions. We report the performance of causal discovery algorithms in generating realistic data in Table~\ref{table:RQ1}.

\begin{figure*}[!tb]
 \begin{minipage}{0.49\textwidth}
  \centering
  \includegraphics[width=1.0\textwidth]{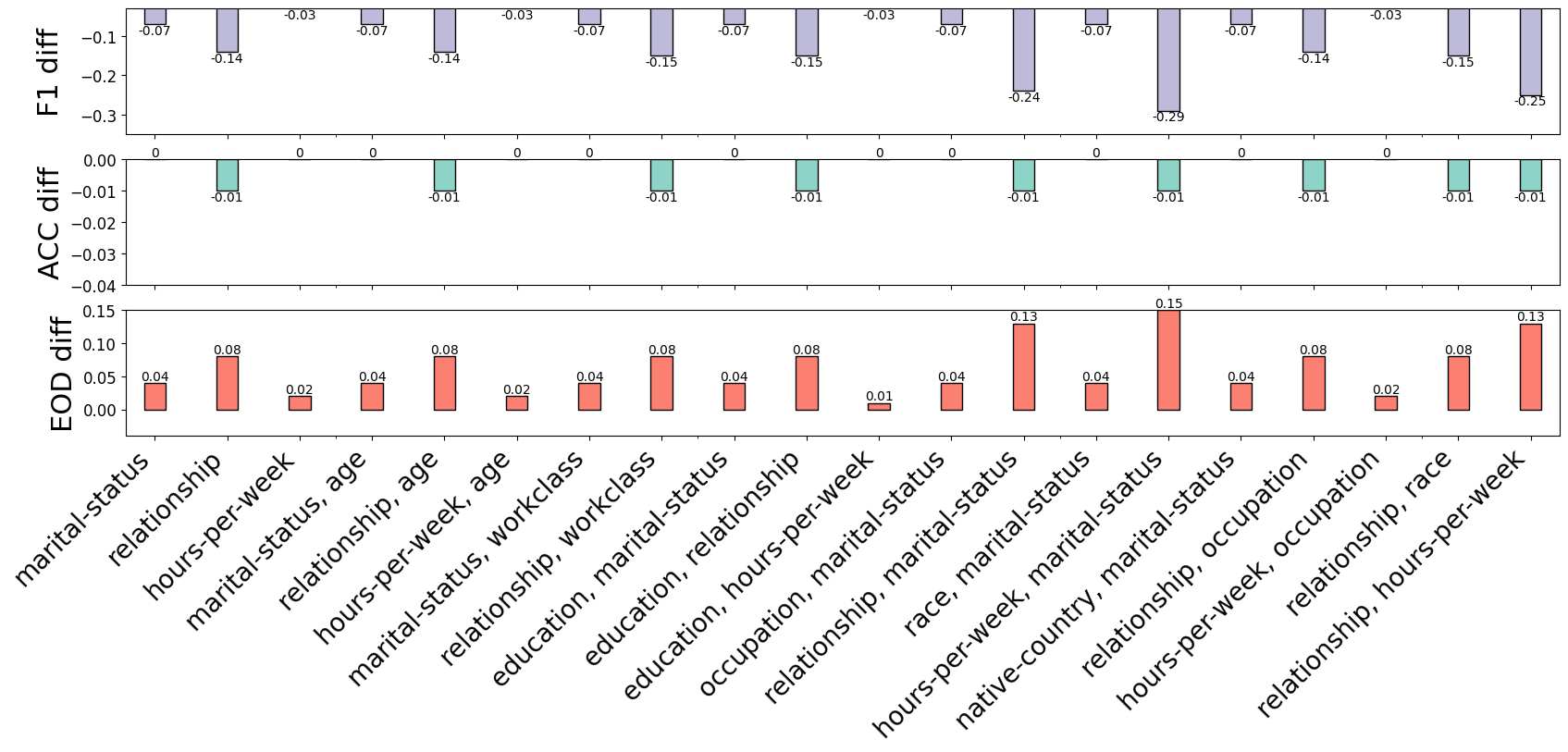}
    \captionsetup{font=footnotesize}
    \caption{Result of EOD on the causal graph~\ref{fig:Adult_DAGs} (b).}    \label{fig:base_results}
 \end{minipage}%
 \vspace{0.25 em}
 \begin{minipage}{0.48\textwidth}
  \centering
  \includegraphics[width=1.0\textwidth]{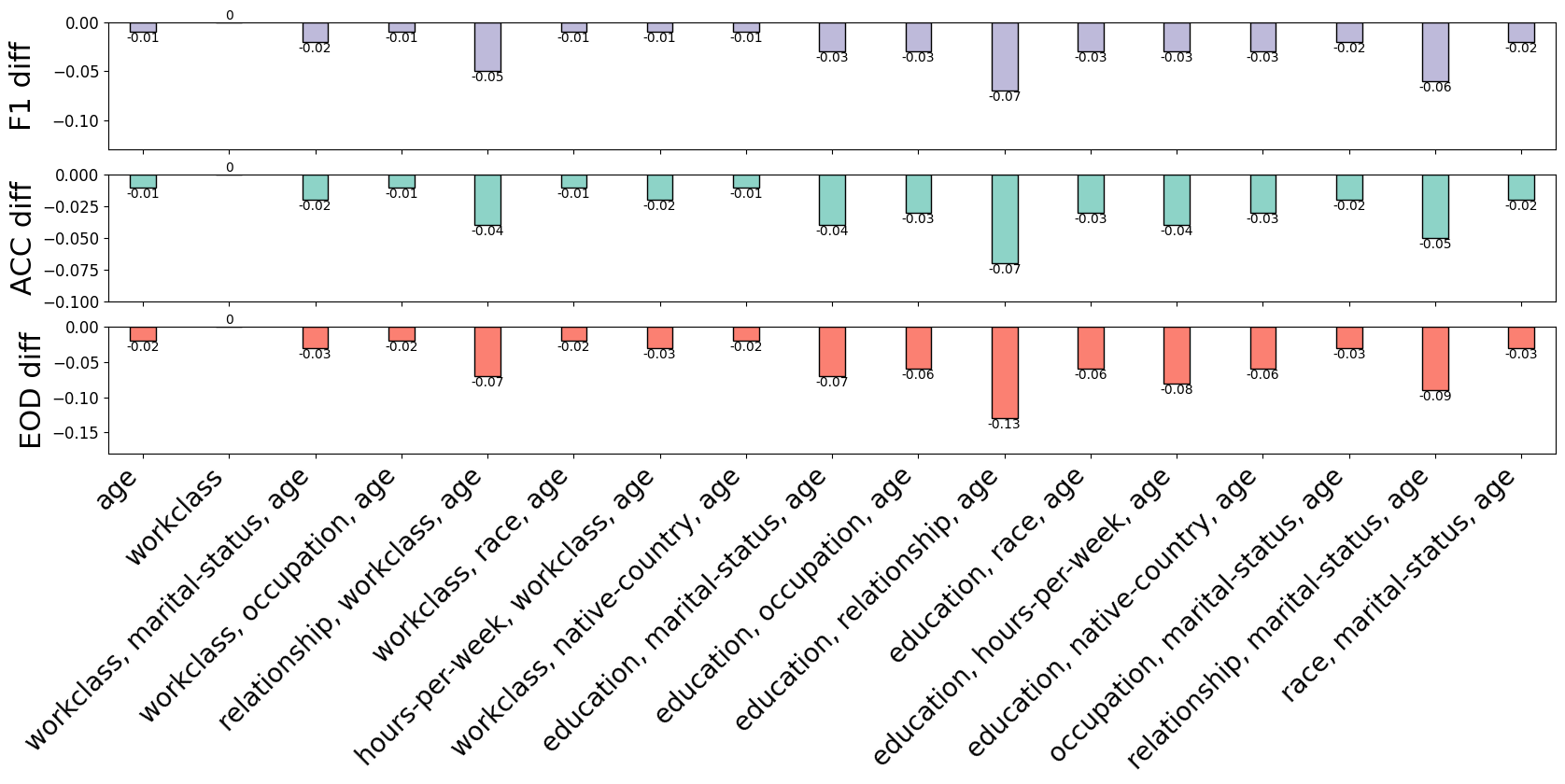}
    \caption{Result of EOD on the causal graph~\ref{fig:Adult_DAGs} (c).}
    \captionsetup{font=footnotesize}
    \label{fig:perturbed_results}
 \end{minipage}
 \vspace{-1.0em}
\end{figure*}

\vspace{0.25 em}
\noindent \textbf{Fairness Practice: Including All Features During Training.} 
% Having discovered the significant role of the causal graph in dropping sensitive features during training and fairness during the first part of our experiment, we are motivated to further explore the impact of dropping non-sensitive attributes on fairness. 
We systematically investigate how selecting features via methods like feature importance influences fairness.

\noindent \textit{Dropping feature randomly.} 
For the causal in Figure~\ref{fig:Adult_DAGs} (b), we train the logistic regression models by excluding different sets of non-sensitive features.
Figure~\ref{fig:base_results} shows that dropping non-sensitive features likely increases the EOD bias. For example, dropping the hours-per-week and marital status feature increased the EOD by
\begin{wrapfigure}{r}{0.25\textwidth}
\begin{flushright}
    \centering
    \includegraphics[width=0.24\textwidth]{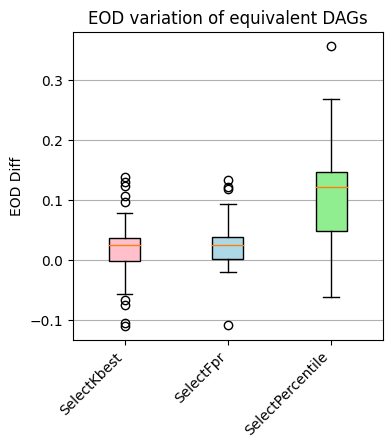}
    \captionsetup{font=small}
    \caption{Selection Operators.}
    \label{fig:Adult_feature_selection_nonsens}
\end{flushright}
\end{wrapfigure}
0.15, while excluding education, relationship, and age decreased the EOD by up to 0.13. These findings aligned with prior research~\cite{zhang2021ignorance,10.1145/3468264.3468536} that indicated an increase in the EOD when non-sensitive attributes are dropped. However, let us consider the equivalence graph in Figure~\ref{fig:Adult_DAGs}  (c).
We repeat the same experiment of dropping non-sensitive features on this neighbor causal graph.
Figure~\ref{fig:perturbed_results} shows the results that are significantly different than the pattern in Figure~\ref{fig:base_results}. Dropping different features consistently decreased the EOD for all cases, with the potential to reduce the EOD (i.e., mitigating bias) by up to 0.13.
% To provide intuitive explanations,
% \vmsays{for Adult}we focus on the car\_type variable that has significant positive and negative
% impacts in the two graphs. In the base graph, there is a backdoor path from race variable
% to car\_type via owned\_redcar (a collider). Therefore, excluding car\_type opens a spurious
% path from race to the outcome variable that increases EOD bias (while including car\_type
% closes the path). In the perturbed graph, our algorithm removed the edge between race and owned\_redcar,
% and it effectively eliminates the backdoor path. Therefore, logistic regression is forced to learn a
% classifier without any spurious correlations with the sensitive attribute race. 
%\gtsays{Maybe explain why this is the case a bit. My understanding is because the correlation between sensitive and non-sensitive features.} \stsays{I provided some explanations. check it out!}

\noindent \textit{Dropping feature via standard feature selection methods.} 
We consider three prevalent feature selection techniques in the scikit-learn library:
\texttt{SelectKBest}~\cite{KBest} (selecting the top K features), \texttt{SelectFpr} \cite{SelectFpr} (selecting the top features based on false positive rates), 
and \texttt{SelectPercentile}~\cite{SelectPercentile}
(selecting the top features based on the percentile). 
Previous research~\cite{10.1145/3468264.3468536} suggests that applying
\texttt{SelectKBest} and \texttt{SelectPercentile} increased unfairness, whereas \texttt{SelectFpr}
did not impact fairness. 
% As before, we use the base causal graph depicted in Figure~\ref{fig:Adult_DAGs} (b), train logistic regression,
% and measure EOD after applying these operators. Then, we apply our search algorithm over equivalence classes and identify DAGs that negate the empirical observations. 
% After training the model, we record and analyze its EOD. Having obtained the EOD for the base causal graph, we initiate our causal perturbation algorithm. This algorithm aims to identify perturbations of the base causal graph (Figure~\ref{fig:insurance_base}) such that when running the three feature selection methods, the EOD demonstrates a decrease compared to the base causal graph. 
Figure~\ref{fig:Adult_feature_selection_nonsens} shows the differences in the EOD between 
graphs \ref{fig:Adult_DAGs} (b) and (c). We found that
the observations for \texttt{SelectKBest} and \texttt{SelectPercentile} hold for the Adult,
but \texttt{SelectFpr} can also degrade fairness. Besides, applying \texttt{SelectKBest} and \texttt{SelectPercentile} can occasionally improve fairness whereas \texttt{SelectFpr} consistently degrades fairness.

\section{Robust Fairness Design For Software}
\label{sec:problem}
In this section, we formalize the notion of local robustness required for developing our empirical results.
% that recommend
% fair designs of the ML training process.
% (e.g., pre-processing over the input features and configuring hyperparameters).

\vspace{0.25 em}
\noindent \textbf{The Dataset.}
Let $\mathcal{F}$ denote the set of all possible features for our dataset. For any feature $f \in \mathcal{F}$, let $\pi_f$ denote the feature space of $f$; for any $A \subseteq \mathcal{F}$ let $\pi_A := \prod_{f \in \mathcal{A}} \pi_f$ be the feature space for the feature set $A$. Additionally assume that the feature space $\mathcal{F}$ has a designated sensitive feature $\hat{f}$ which has a boolean feature space i.e. $\pi_{\hat{f}} = \{ 0, 1 \}$. Let $Y = \{ 0, 1 \}$ represent our boolean output space where $1$ denotes a favorable outcome, and $0$ denotes an unfavorable outcome. Let $\mathbb{D}$ represent the class of all datasets that can be constructed from $\pi_{\mathcal{F}} \times Y$. Then any $\mathcal{D} \in \mathbb{D}$ is a set of samples, written as $(\mathbf{\vec{x_i}}, y_i)_{i}$, where $\mathbf{\vec{x_i}} \in \pi_{\mathcal{F}}$ are feature vectors and $y_i \in \{ 0, 1 \}$ are boolean output variables.  Additionally, for any $A \subseteq \mathcal{F}$, let $\mathcal{D}_{A}$ represent the reduced dataset $(\mathbf{\vec{x^A_i}}, y_i)_{i}$, where forall $i$, $\mathbf{\vec{x^A_i}}$ is the vector $\mathbf{\vec{x_i}}$ restricted to the feature set $A$. Datasets can be generated from generative models such as structural causal models~\cite{pearl2018book}.

\begin{comment}
\textbf{The Causal Graph.}
 We first compute a special type of causal graph called a bayesian network (BN) from $\mathcal{D}$. A bayesian network for $\mathcal{D}$ is a dag $\mathcal{G}$ on all the features of $\mathcal{D}$ together with a set of random variables $\{ X_v\}_{v=1}^m$ for each feature $v$ which satisfy the Markov condition i.e. for any events $B_1, B_2, .. B_m$ the joint distribution of all random variables can be factorized as:
\begin{align*}
    \probP{}(X_1 \in B_1, X_2 \in B_2, .. X_m \in B_m) = \prod_{v = 1}^m \probP{}(X_v \in B_v | pa(X_v))    
\end{align*}
where $pa(X_v)$ denotes the parents of node $X_v$. $\mathcal{G}$ can be computed from $\mathcal{D}$ using the causal discovery algorithms in \cite{binkyte2022causal}.
\end{comment}

\vspace{0.25 em}
\noindent \textbf{The ML Paradigm.} Any given ML algorithm $\mathcal{S}$ (e.g., logistic regression) allows a set of hyperparameters $\mathcal{H}_{\mathcal{S}}$. For any $h \in \mathcal{H}_{\mathcal{S}}$, let $\Psi_h$ be the hyperparameter space of $h$, and let $\Psi_{\mathcal{S}} := \prod_{h \in \mathcal{H}_{\mathcal{S}}} \Psi_h$ be the complete hyperparameter space. Then, we define \textit{the parameter set} of training process as $\mathcal{H}_{\mathcal{S}} \times 2^{\mathcal{F}}$ with its corresponding parameter space defined as $\Theta_{\mathcal{S}} := \Psi_{\mathcal{S}} \times 2^{\mathcal{F}}$. Given a dataset $\mathcal{D} \in \mathbb{D}$ and a parameter configuration $(\theta, A) \in \Theta_{\mathcal{S}}$, a ML model for $\mathcal{S}$ learns a function $M: \pi_{A} \rightarrow \{ 0, 1 \}$ by using the reduced dataset $\mathcal{D}_{A}$ and hyperparameter configuration $\theta$ to learn the unknown weights. 
The fitness of a ML model is measured through the accuracy or F1-score of the function $M$ learned w.r.t a validation dataset $\mathcal{D}^* \in \mathbb{D}$. The accuracy of $M$ w.r.t $\mathcal{D}^*$, denoted $ACC^M$, is defined as the ratio of correct results on $\mathcal{D}^*$ to the total number of samples. In order to define the F1 score of $M$, we need to first define the precision and recall of $M$ w.r.t $\mathcal{D}^*$. The precision of $M$ w.r.t $\mathcal{D}^*$, denoted $Prec^M$, is defined as the ratio of correctly predicted favorable outcomes to total predicted favorable outcomes, whereas the recall of $M$ w.r.t $\mathcal{D}^*$, denoted $Rec^M$, is defined as the ratio of correctly predicted favorable outcomes to total favorable outcomes. The F1 score of $M$ w.r.t $\mathcal{D}^*$, denoted $F1^M$, is the harmonic mean of $Prec^M$ and $Rec^M$.

\vspace{0.25 em}
\noindent \textbf{Fairness of ML model.} The fairness of an ML model for a given dataset $\mathcal{D}$, feature set $A$, and hyperparameter configuration $\theta$ is analyzed by studying the bias of the function $M$ learnt with respect to $\hat{f} \in \mathcal{F}$. We first define the true positive rate of our learned function $M$ conditioned to the event that feature $\hat{f}$ has value $b \in \{ 0, 1 \}$, denoted by $TPR^{M}(b)$, and defined by the following formula:
$$
TPR^{M}(b) = \frac{|\{ (\mathbf{\vec{x_j}}, y_j) {\in}{\mathcal{D}}_A \: : \: x_j({\hat{f}}){=}b, M(\mathbf{\vec{x_j}}){=}1, y_j{=}1 \} |}{|\{ (\mathbf{\vec{x_j}}, y_j)\in \mathcal{D} \: : \: x_j({\hat{f}}) = b\}|}
$$
Using this, we can define the bias of $M$ w.r.t sensitive feature $\hat{f}$ using the equal opportunity difference $(EOD)$ metric. The EOD of $M$ w.r.t a sensitive feature $\hat{f}$ is defined as:
$$EOD^{M} = |TPR^{M}(1) - TPR^{M}(0)| $$
As any learning on  $\mathcal{S}$ can be viewed as using $\mathbb{D} \times \Theta_{\mathcal{S}}$ to produce a function $M :  \pi_A \rightarrow \{0,1\}$ and each such function $M$ learnt has an associated bias value $EOD^{M}$, we can view the bias of an ML model for $\mathcal{S}$ as a function from $\mathbb{D} \times \Theta_{\mathcal{S}}$ to $[0,1]$ defined as the EOD value for the learned ML function $M$, where $M$ is learned via the training process with a dataset from $\mathbb{D}$ and parameter configuration from $\Theta_{\mathcal{S}}$ with an acceptable $F1^M$ and $ACC^M$. Let us call this function $\textit{bias}_{\mathcal{S}}: \mathbb{D} \times \Theta_{\mathcal{S}} \rightarrow [0,1]$.  

\vspace{0.25 em}
\noindent \textbf{Problem Definition.} 
We validate local robustness in existing fairness properties on social-critical datasets and specific parameter configurations. We first restrict $\mathbb{D}$ to denote only datasets which may appear in the real-world\footnote{We ensure this by generating our datasets from causal graphs which in turn have been derived from real-world datasets only. We describe this dataset generation process in more detail later.}. Given a `real-world' dataset $\mathcal{D} \in \mathbb{D}$ and two parameter configurations, $\theta_1, \theta_2 \in \Theta_{\mathcal{S}}$, a fairness property is a first order formula with parameters $\mathcal{D}, \theta_1$ and $\theta_2$, denoted as $prop(\mathcal{D}, \theta_1, \theta_2)$. Given a neighborhood relation on `real-world' datasets $\mathbb{D}$, denoted $\sim$, which captures how `similar' two datasets are, we wish to answer the following research question:  
\noindent \emph{Given a `real-world' dataset $\mathcal{D}$ and configurations $\theta_1$ and $\theta_2$ that satisfies
a fairness design property $prop(\mathcal{D}, \theta_1, \theta_2)$,
the research problem is to find a `real-world' dataset $\mathcal{D}' \sim \mathcal{D}$,
s.t. $prop(\mathcal{D}', \theta_1, \theta_2)$ fails.}
% \end{tcolorbox}

For example, an existing property is that dropping sensitive features from the default configuration for a ML-algorithm $\mathcal{S}$ increases fairness. Given a default hyperparameter $h_0$, such a property, called $dropSen(\mathcal{D}, \theta_1, \theta_2)$, can be defined as: 
\begin{align*}
dropSen(\mathcal{D}, \theta_1, \theta_2) &{\equiv} \Big((\exists h_0) \:\: \theta_1 {=} ({h_0}, \mathcal{F} ) {\land} \theta_2 = (h_0, \mathcal{F} {\setminus} \hat{f}) \Big) \\
&\land (\textit{bias}_{\mathcal{S}}(\mathcal{D}, \theta_1) > \textit{bias}_{\mathcal{S}}(\mathcal{D}, \theta_2))    
\end{align*}
The $dropSens$ is not locally robust if for some `real-world' dataset $\mathcal{D}' \sim \mathcal{D}$ and parameter configurations $\theta_1,\theta_2$, $dropSens(\mathcal{D}, \theta_1, \theta_2)$ holds \texttt{true}, but $dropSens(\mathcal{D}', \theta_1, \theta_2)$ holds \texttt{false}.

\section{Approach}
\label{sec:approach}

\vspace{0.25em}
Our approach consists of the following phases: a) partial causal graph discovery, b) inferring structural causal models (SCM) to generate datasets, and c) search over the SCMs to validate the robustness of fairness practices. 

\noindent \textbf{A. Partial causal graph discovery.} We first utilize three causal discovery algorithms~\cite{spirtes2000causation,chickering2002optimal,10.5555/3020419.3020473} to infer the direction of edges between features (i.e., cause-effect relation). We use the PC algorithm~\cite{spirtes1991algorithm}, GES algorithm~\cite{chickering2002optimal}, and SIMY algorithm~\cite{10.5555/3020419.3020473} to infer the direction of edges between features. However, these algorithms often infer the Completed Partially Directed Acyclic Graph (CPDAG)
where the directions of some edges have not been resolved. We consider each directed acyclic graph (DAG) as an
equivalence graph. 

\vspace{0.25em}
\noindent \textbf{B. Inferring structural causal model.}
In our study, we implement Bayesian inference methods using the STAN probabilistic programming language to estimate the weights of edges in causal graphs (i.e., DAG), specifically focusing on determining the strengths of the relationships between various variables. Our approach involves assigning appropriate distributions to different types of variables---continuous, discrete, and Boolean---based on their characteristics. These distributions are then encoded as probabilistic models in STAN to infer posterior distributions for the edge weights using MCMC algorithms. Our methodology follows the principles in ~\cite{Counterfactual-Fairness}.

To provide a clearer illustration, let's consider specific examples from the causal graph Figure~\ref{fig:Adult_DAGs} (b). For a continuous variable like hr, we used a Gaussian distribution modeled as 
$\text{hr} \sim \mathcal{N} (b_{\text{hr}}  +  w^o_{\text{hr}} o  +  w^e_{\text{hr}} e  +  w^{\text{gender}}_{\text{hr}} gender,  \sigma_{\text{hr}})$, where $b_{\text{hr}}$ is the bias term, the weights 
$w^o_{\text{hr}}$, $w^e_{\text{hr}}$, and $w^{\text{gender}}_{\text{hr}}$ correspond to the influence from other variables, and $\sigma_{\text{hr}}$ is the standard deviation.  For discrete variables like age, we employ a Poisson distribution, represented as $\text{age} \sim \text{Poisson}(\exp(b_{\text{age}} + w^{\text{gender}}_{\text{age}} gender + w^o_{\text{age}} o + w^e_{\text{age}} e))$, where each term incorporates the impact of different variables. Finally, for Boolean variables like e, we utilize a Bernoulli distribution, as in  $\text{e} \sim \text{Bernoulli} (b_e  +  w^o_e o  +  w^{\text{gender}}_e gender)$. After inferring the weights of the causal graphs, we focus on generating samples from the posterior distributions of these DAGs.

\vspace{0.25em}
\noindent \textit{In-distribution data generation.} The next step of our approach is to generate in-distribution neighbor datasets. While traditional methods like GANs~\cite{xu2019modeling}, VAEs~\cite{8285168}, and bootstrapping~\cite{1609588} can generate in-distribution data, they fall short when it comes to creating neighbor datasets. 
% with controlled and nuanced changes for evaluating the robustness of fairness practices in ML systems. 
GANs~\cite{xu2019modeling,math11040977,10.1007/978-3-031-35891-3_26, pmlr-v157-zhao21a} and VAEs~\cite{Zhang2015LearningCF,10230343,hajighasemi2023multimodal,8285168,ISLAM2021105950,vora2023real}, despite their ability to generate realistic synthetic data, lack the transparency and control needed to understand and manipulate feature relationships, limiting their use in creating datasets with specific variations. Similarly, bootstrapping~\cite{647043,10.1162/089976600300015204}, while effective for generating in-distribution data, does not provide insights into the conditions under which these neighboring datasets are obtained or allow for deliberate manipulation of the type of distribution shift applied. In contrast, our approach leverages causal graphs, which offer a more transparent and controllable method for generating neighbor datasets. By explicitly representing feature relationships and their directionality, causal graphs enable the creation of datasets with predetermined variations, allowing us to systematically explore how fairness design practices behave under different conditions and identify scenarios where the empirical findings as a fairness property might differ maximally. 

To ensure that generated samples by the causal graphs match the actual data distribution (in-distribution data generations), we cluster the original training dataset and set a threshold based on the average Euclidean distance to the centroids of these clusters (validated over some validation dataset). We then evaluate each generated sample from a causal discovery algorithm against this distance threshold. The effectiveness of each algorithm is measured by its success rate in generating samples that meet this distance criterion, establishing in-distribution samples.

\vspace{0.25em}
\noindent \textit{Causal inference under distribution shifts.} So far, our analysis assumes in-distribution neighborhood datasets. To evaluate the local robustness of the best practices, we also analyze them under distribution
shifts. There are three primary distribution shifts: prior probability shift, covariate shift, and concept drift~\cite{kull2014patterns,varshney2021trustworthy}. In this paper, we only consider prior probability shift
(also known as label shift), where the label distributions are different between two populations and their samples.
For example, the percentage of samples with incomes over \$50K is 30\% and 39\% in the US Adult census data of 2015 and 2016, respectively. 
To imitate the prior probability shift
during the causal inference, we add a constant term to the bias term of the label feature and search the space
of this term to generate (out-of-distribution) datasets with a prior probability shift. 

\vspace{0.25em}
\noindent \textbf{C. Search to validate fairness practices.}
Given an input dataset, a search space (i.e. equivalence causal graphs for generating in-distribution and out-of-distribution samples), and a property derived from a practice in the fair ML software development; we utilize a search algorithm to identify two `similar' datasets generated from two equivalence DAGs, where one satisfies the property, but the other one does not. We train logistic regressions over the training datasets and evaluate their performance (i.e., accuracy, F1) and fairness (i.e., EOD) over the test data.
We record the causal graphs that manifest the maximum observed difference in fairness,
and leverage the weights of those graphs for the next round of search. 
If we find two causal graphs that contradict in satisfying the property during the search, we terminate the search and return the identified graphs. Otherwise, we stop the search after a timeout. 

\vspace{0.5em}
\noindent \textbf{Putting everything together.}  
Algorithm~1 (see appendix) describes our approach to investigate the relationship between the causal graphs and common
practices in fairness training of ML models. We first use the input dataset to obtain the causal graph skeleton (CPDAG). Then, we generate all possible equivalence DAGs from each CPDAG and infer a set of 1,000 causal graphs for each DAG with slightly different models. We then generate a data sample
and validate it by comparing its Euclidean distance to the closest centroid of 100 clusters formed over the training dataset. We accept the sample only if it falls within the average distance calculated previously over the validation dataset. This criterion is also used to evaluate the performance of different causal discovery algorithms. 

Once we identify a set of causal graphs, we run the search algorithm to validate whether a fairness practice (i.e., property)
holds true between two similar datasets. We note that the search depends on the type of property. For example,
\emph{if the type of analysis is the effect of excluding sensitive attributes on fairness}, we simply exclude sensitive attributes from the generated data samples during training and measure the EOD bias. On the other hand, \emph{if the type of analysis is the feature selection}, we use the following methods: \texttt{random} (i.e., exclude a subset of features at random up to 3 features during training), \texttt{SelectKBest}~\cite{KBest} (i.e., only include top $K$ features in training), \texttt{SelectFpr}~\cite{SelectFpr} (i.e., include features based on false positive rates), and \texttt{SelectPercentile}~\cite{SelectPercentile} (i.e., select top features based on their percentile scores) to select a subset of feature for training.
We guide the search based on the most promising pair of equivalence graphs that have witnessed the largest bias differences.

\section{Experiments}
\label{sec:experiments}

{\footnotesize
\begin{table}[!t]
\caption{Datasets used in our experiments.}
\centering
\resizebox{0.49\textwidth}{!}{
\begin{tabular}{|l|l|l|l|ll|ll|}
  \hline
  \multirow{2}{*}{\textbf{Dataset}} & \multirow{2}{*}{\textbf{\#Instances}} & \multirow{2}{*}{\textbf{\#Features}} &  \multirow{2}{*}{\textbf{Prot. Att}} & \multicolumn{2}{c|}{\textbf{Dist Accuracy}}&\multicolumn{2}{c|}{\textbf{Outcome Label}} \\
  &&  && TPR & FNR & \textit{Label 1} & \textit{Label 0}  \\
  \hline
  Adult~\cite{Adult} & 48,842 & 10 & Sex & 0.95&0.11& Income $\geq$ 50K & Income $<$ 50K \\ 
  \hline
  Compas~\cite{compas-dataset} & 7,214 & 6 & Race & 0.94&0.12 & Not Reoffend & Reoffend \\ 
  \hline
  
  Bank~\cite{Dua:2019-bank} &45,211  & 16 & age & 0.98& 0.17& Subscriber & Non-Subscriber\\ 
  \hline
  Law School~\cite{Counterfactual-Fairness} & 21,791 & 4 & Sex & 0.96&0.34 & $1$-year Succeed & $1$-year Failed\\ 
  \hline
  Student~\cite{Student-performance} & 1,044 & 17 & Sex & 0.93& 0.36& Passed & Not passed\\ 
  \hline
  Heart Disease~\cite{Heart-disease} & 297 & 9 & Sex & 0.92& 0.31& Disease & Not Disease\\ 
  \hline
\end{tabular}
}
\vspace{-1.0 em}
\label{table:dataset}
\end{table}
}

% {\footnotesize
\begin{table}[!b]
\caption{Effectiveness of causal discovery algorithms.}
\centering
\resizebox{0.48\textwidth}{!}{
\begin{tabular}{|l|l|c|cccc|c|}
  \hline
  \multirow{2}{*}{\textbf{Dataset}} & \multirow{2}{*}{\textbf{Algorithm}} & \multirow{2}{*}{\textbf{\#DAGs}} &  \multicolumn{4}{c|}{\textbf{Succ rate}}  &  \multirow{2}{*}{\textbf{Dist}} \\
   % &&&& &&
    & & & $Avg$&$Std$&$Min$&$Max$& \\
  \hline
   &PC&$32$&$0.4$ & $0.02$ & $0.38$ & $0.43$ & $2.9$  \\
   &\cellcolor{gray!35} GES&\cellcolor{gray!35}$4$& \cellcolor{gray!35}$0.46$ &\cellcolor{gray!35} $0.19$ & \cellcolor{gray!35}$0.13$ & \cellcolor{gray!35}$0.58$ & \cellcolor{gray!35} $3.4$ \\
   Adult~\cite{Adult} &\cellcolor{gray!35} SIMY&\cellcolor{gray!35} $4$&\cellcolor{gray!35} $0.53$ & \cellcolor{gray!35} $0.04$ & \cellcolor{gray!35} $0.48$ & \cellcolor{gray!35} $0.57$ & \cellcolor{gray!35} $3.1$ \\
   &RND&$40$&$0.0$ & $0.00$ & $0.0$ & $0.01$ & $6.9$  \\
   &EQ&$40$&$0.02$ & $0.03$ & $0.0$ & $0.12$ & $6.6$    \\

  \hline
 
    &\cellcolor{gray!35} PC&\cellcolor{gray!35} $16$&\cellcolor{gray!35} $0.59$ & \cellcolor{gray!35} $0.07$ & \cellcolor{gray!35} $0.51$ & \cellcolor{gray!35} $0.7$ & \cellcolor{gray!35} $0.4$\\ 
  &\cellcolor{gray!35} GES&\cellcolor{gray!35} $8$&\cellcolor{gray!35} $0.61$ & \cellcolor{gray!35} $0.08$ &\cellcolor{gray!35}  $0.53$ & \cellcolor{gray!35} $0.7$ & \cellcolor{gray!35} $0.5$\\
  Compas~\cite{compas-dataset}&SIMY&$16$&$0.55$ & $0.0$ & $0.54$ & $0.55$ & $0.5$\\
   &RND&$30$&$0.10$ & $0.05$ & $0.045$ & $0.23$ & $2.7$  \\
   &EQ&$30$&$0.11$ & $0.08$ & $0.04$ & $0.34$ & $2.8$ \\
  \hline

    &PC&$2$&$0.26$ & $0.0$ & $0.26$ & $0.26$ & $8.0$ \\
  &\cellcolor{gray!35} GES&\cellcolor{gray!35} $12$&\cellcolor{gray!35} $0.31$ &\cellcolor{gray!35}  $0.2$ & \cellcolor{gray!35} $0.01$ & \cellcolor{gray!35} $0.57$ & \cellcolor{gray!35} $6.6$ \\
  Bank~\cite{Dua:2019-bank}&SIMY&$16$&$0.1$ & $0.04$ & $0.05$ & $0.14$ & $6.8$ \\
   &RND&$30$&$0.0$ & $0.0$ & $0.0$ & $0.01$ & $10.0$ \\
   &EQ&$30$&$0.0$ & $0.0$ & $0.0$ & $0.02$ & $9.5$\\
  \hline

    &PC&$8$&$0.0$&0.0& $0.0$& $0.0$& $NA$ \\
   &\cellcolor{gray!35}GES&\cellcolor{gray!35}$18$&\cellcolor{gray!35}$0.65$ & \cellcolor{gray!35}$0.0$ & \cellcolor{gray!35}$0.64$ & \cellcolor{gray!35}$0.65$ & \cellcolor{gray!35}$0.4$ \\
  Law School~\cite{Counterfactual-Fairness}&\cellcolor{gray!35}SIMY&\cellcolor{gray!35}$18$&\cellcolor{gray!35}$0.65$ & \cellcolor{gray!35}$0.01$ & \cellcolor{gray!35}$0.64$ & \cellcolor{gray!35}$0.66$ & \cellcolor{gray!35}$0.4$ \\
   &RND&$44$&$0.0$ & $0.02$ & $0.0$ & $0.09$ & $12.3$  \\
   &EQ&$44$&$0.0$ & $0.01$ & $0.0$ & $0.03$ & $13.2$ \\
  \hline
  
    &\cellcolor{gray!35}PC&\cellcolor{gray!35}$1$&\cellcolor{gray!35}$0.28$ & \cellcolor{gray!35}$0.0$ & \cellcolor{gray!35}$0.28$ & \cellcolor{gray!35}$0.28$ & \cellcolor{gray!35}$3.5$ \\
  &GES&$4$& $0.19$ & $0.0$ & $0.19$ & $0.2$ & $3.6$ \\
  Students~\cite{Student-performance}&\cellcolor{gray!35}SIMY&\cellcolor{gray!35}$1$& \cellcolor{gray!35}$0.23$ & \cellcolor{gray!35}$0.0$ & \cellcolor{gray!35}$0.23$ & \cellcolor{gray!35}$0.23$ & \cellcolor{gray!35}$3.4$\\
  &RND&$6$&$0.0$ & $0.0$ & $0.0$ & $0.0$ & $14.9$  \\
  &EQ&$6$&$0.0$ & $0.0$ & $0.0$ & $0.0$ & $16.9$ \\
  \hline
    &\cellcolor{gray!35}PC&\cellcolor{gray!35}$2$& \cellcolor{gray!35}$0.17$ & \cellcolor{gray!35}$0.07$ & \cellcolor{gray!35}$0.1$ & \cellcolor{gray!35}$0.24$ & \cellcolor{gray!35}$2.2$ \\
  &\cellcolor{gray!35}GES&\cellcolor{gray!35}$8$&\cellcolor{gray!35}$0.17$ & \cellcolor{gray!35}$0.05$ & \cellcolor{gray!35}$0.08$ & \cellcolor{gray!35}$0.23$ & \cellcolor{gray!35}$2.1$ \\
  Heart~\cite{Heart-disease}&\cellcolor{gray!35}SIMY&\cellcolor{gray!35}$8$&\cellcolor{gray!35}$0.18$ & \cellcolor{gray!35}$0.06$ & \cellcolor{gray!35}$0.08$ & \cellcolor{gray!35}$0.24$ & \cellcolor{gray!35}$2.1$\\
 &RND&$18$&$0.0$ & $0.0$ & $0.0$ & $0.0$ & $66.4$  \\
 &EQ&$18$&$0.0$ & $0.0$ & $0.0$ & $0.0$ & $70.5$ \\
  \hline
\end{tabular}
}
\label{table:RQ1}
\label{table:RQ1_old}
\end{table}

% \begin{figure}[t]
%     \includegraphics[width=0.49\textwidth]{Figures/adult_base_cg.png}
%     \caption{Base causal graph for Adult dataset.}
%     \label{fig:Adult_base}
%     \centering
% \end{figure}

\noindent \textbf{Datasets and machine learning model.} 
We utilize six commonly used datasets from the fairness literature~\cite{Counterfactual-Fairness,chakraborty2020fairway,10.1145/3510003.3510202}. Table~\ref{table:dataset}
describes the properties of these datasets. To assess the efficacy of our distance function in identifying in-distribution data samples, we performed a split of each dataset into training and validation sets. We computed the average distance criteria on the training set and evaluated the performance on the validation set by measuring the True Positive Rate (TPR). To further understand the behavior of False Negative Rate (FNR), we generated a random uniform test set and applied the distance function to it. The results of these tests are reported in the Dist. Accuracy with TPR and FNR columns of Table~\ref{table:dataset}. The high TPR combined with a low FNR indicates the reliability of our distance criterion in evaluating the accuracy of generated samples by the causal graph algorithms. Besides the datasets, we utilize the logistic regression (LR), decision tree (DT), and support vector machine (SVM) algorithms from the scikit-learn library to infer the ML models throughout this paper. 

\noindent \textbf{Technical Details.}
We implement our tool with TensorFlow v2.10.0, scikit-learn v1.2.2, Rstan v2.32.3, and pcalg v2.7.9. We run all the experiments
on an Ubuntu 20.04.4 LTS OS sever with AMD Ryzen Threadripper PRO 3955WX 3.9GHz 32-cores X  CPU and two NVIDIA GeForce RTX 3090 GPUs. We set $4$ hours and $0.05$ for timeout and the accuracy/F1 loss tolerance, respectively. The inference of posterior distributions for DAGs depends on the number of features and the dataset size. In our experiments, it takes an average of four hours per dataset. However, this is a one-time computational cost and does not affect the search time. Our search algorithm efficiently identifies two contradicting causal graphs in about $5$ minutes.
When analyzing hyperparameters, we adopt the evolutionary search algorithm introduced by Tizpaz-Niari et al. \cite{10.1145/3510003.3510202}, which uses mutation operators to explore the ML model hyperparameter space and identify configurations that minimize fairness violations. In our study, we treat this tool as a fairness intervention without modifying its internal logic. We execute the tool for $4$ hours per dataset to extract fair configurations and then use our causal framework to test the robustness of these configurations across neighboring datasets.
We repeat our experiments $30$ times and employ the Scott-Knott statistical significance test~\cite{10.1109/ICSE48619.2023.00133} to validate our results (higher rank values are reported with bold fonts). 

\noindent \textbf{Design Choices.}
We generate 1,000 causal graphs per equivalence class to ensure sufficient structural diversity in representing neighborhood datasets, following established practices in causal modeling \cite{Counterfactual-Fairness}. To validate in-distribution sampling, we apply k-means clustering with 100 clusters—a value chosen empirically to offer adequate granularity while remaining computationally feasible. Importantly, this number of clusters acts as a tunable hyperparameter that can be adjusted based on the characteristics of the dataset.

\noindent \textbf{Research Questions.}
Here are four research questions:

\begin{enumerate}[start=1,label={\bfseries RQ\arabic*},leftmargin=3em]
\item What is the quality of data generation by different causal discovery algorithms?
% Does training unaware models of sensitive features always leads to have unfair models? What is the impact of the causal network of the dataset when it comes to unaware models?

\item Are the best fairness practices robust when non-sensitive or sensitive attributes are dropped during training with neighborhood causal graphs?
%Are feature selection practices for fairness robust? 
% What is the impact of the underlying causal graph on the fairness of machine learning models when non-sensitive attributes are dropped during training?

\item Do hyperparameter configurations remain robust w.r.t fairness of outcomes when the underlying causal representations slightly change?

\item Are the post-processing bias mitigation practices locally robust? 
% \stsays{update this!}
\end{enumerate}
% \noindent \textbf{Research Questions.}
% In this study, we aim to answer the following $4$ research questions.

% \begin{enumerate}[start=1,label={\bfseries RQ\arabic*},leftmargin=3em]
% \item Which causal discovery algorithms can generate a realistic dataset (representing
% the underlying distributions of different protected attributes)
% and preserve the underlying discrimination based on metrics like demographic parity, etc.?

% \item 
% Does excluding sensitive attributes always degrade fairness? How to systematically study
% the influence of training with sensitive attributes over fairness? 
% Does feature selections always degrade fairness? How to systematically understand
% the impact of selecting non-sensitive attributes on fairness? 

% \item Can hyperparameter configurations systematically aggravate or mitigate unfairness?
% How does the training dataset influence optimal hyperparameter selection?

% \item Do the empirical findings hold for out-of-distribution samples? 
% % \stsays{he validity of dropping, excluding features when the samples may be out-of-distributions. The motivation can be the fact that the dataset might be used for a different social context that was intended (e.g., insurance car example in the state of Texas vs. Wyoming!). In doing so, we need to be careful on what accounts for out-of-distribution samples.}
% \end{enumerate}
\noindent All subjects, data, and our tool are publicly accessible: \href{https://anonymous.4open.science/r/Fairness_Practices_Robustness_Testing-E806/}{Link}.

% \subsection{The performance of causal discovery algorithms to generate realistic datasets (RQ1)}
\noindent \textbf{Causal algorithms for the data generation (RQ1).}
\label{sec:experiments:RQ1}
We investigate the effectiveness of three widely-used causal discovery algorithms PC~\cite{spirtes2000causation}, GES~\cite{chickering2002optimal}, and SIMY~\cite{10.5555/3020419.3020473}. We also include two baselines alongside the causal discovery algorithms. The first baseline, Random Weights (RND), assigns random weights from a standard normal distribution $\mathcal{N} (0,1)$ to the edges in the causal graph. The second baseline, Equal Weights (EQ), assigns equal weights from $\mathcal{N} (0,1)$ to all edges, creating a uniform structure. These baselines serve as an ablation study mechanism. 
Table~\ref{table:RQ1} presents the results from our experiment. In this table, the column \texttt{\#DAGs} indicates the number of DAGs possible in the CPDAG produced by each algorithm. The column  \texttt{\#Succ rate} shows the percentage of samples generated by each algorithm that met our distance criteria (as introduced for each dataser in Table~\ref{table:dataset}) where the sub-columns \texttt{Avg}, \texttt{Std}, \texttt{Min}, and \texttt{Max}, provide summary statistics of these results. The \texttt{Dist} column details the average distance between the generated samples and their nearest neighbors of the training dataset. In short, these metrics calculate the proportion of accepted samples for each causal graph as its success rate.

First, comparing the success rate of causal algorithms against the baselines RND and EQ, it is evident that causal discovery algorithms play a critical role in generating data samples from the training distribution. The results also suggest that the performance of causal discovery algorithms significantly varies depending on the characteristics of the input dataset. For instance, with the Bank dataset, GES outperforms others with an average success rate higher by 16\% and a maximum of 34\%. Conversely, GES shows the lowest average and maximum success rates in the Student dataset. 
Additionally, the summary statistics of success rates enable us to identify classes of DAGs more likely to generate in-distribution data. This insight also helps in excluding algorithms and their corresponding DAGs that exhibit lower potential during the search phase. For example, in the Adult dataset, while PC shows a success rate of 40\%, the GES and SIMY demonstrate significantly better performance, both on average and at their maximum rates. 
% Hence, we exclude DAGs from the PC algorithm in our search for the Adult dataset.
% Overall, this approach allows us to identify algorithms capable of generating data with up to a 71\% success rate. It also aids in pinpointing algorithms that are less effective in producing realistic data.

\begin{answerbox}
\textbf{Answer RQ1:}
Causal discovery algorithms like PC, GES, and SIMY show varying effectiveness in generating in-distribution data depending on the input dataset. The success rate criterion helped us exclude graphs with a low accuracy for the search. 
\end{answerbox}
% \input{Sections/Table_RQ1_validation}
% \noindent \textit{Validation of the synthetic data.}
% But when considering the standard deviation of the success rates GES has almost $18\%$ standard deviation suggesting the possibility of one or more DAGs in the CPDAG generated by GES that achieved a high success rate. Utilizing the statistics information of success rates enables us to identify the algorithms that could generate realistic data as well as exclude the ones with low performance for our search algorithm in the next studies. Highlighted rows in the table show the algorithm selected to be more likely to generate realistic data based on information obtained from their success rates. 
% \begin{tcolorbox}[boxrule=1pt,left=1pt,right=1pt,top=1pt,bottom=1pt]
% \textbf{Answer RQ1:}
% The results in the Table ~\ref{table:RQ1} suggest that different causal discovery algorithms show different performances depending on the input dataset and utilizing the proposed success rate enables us to identify those algorithms that are more likely to generate realistic data that fit the true distribution of the input dataset. This approach also enables us to exclude the algorithms that show a low performance to narrow the space of the search for other research questions. Overall, our approach could identify algorithms that generate data up to $71\%$ success rate as well as identify algorithms that could not generate real data at all. 
% \end{tcolorbox}

% {\footnotesize
\begin{table*}[!t]
\caption{Sensitive \& Non-Sensitive Feature Selection Methods and Their Local Robustness over Causal Graphs.}
\centering
\resizebox{\textwidth}{!}{%
\begin{tabular}{|c|c|c|c|c|c|c|c|c|c|c|c|c|c|c|c|c|c|c|c|}

\hline
\multirow{2}{*}{\textbf{Model}}&\multirow{2}{2.5em}{\textbf{Dataset}}  &\multicolumn{4}{| c |}{\hfil\textsc{SelectKBest~\cite{KBest}}} & \multicolumn{4}{| c |}{\hfil\textsc{SelectFpr~\cite{SelectFpr}}}  &\multicolumn{4}{| c |} {\hfil\textsc{SelectPercentile~\cite{SelectPercentile}}} & \multicolumn{5}{| c |}{\hfil\textsc{DropSensParam}} \\\cline{3-19}
&&\hfil\texttt{\#Edge diff}&\hfil\texttt{EOD diff}&\hfil\texttt{Acc diff} &\hfil\texttt{F1 diff}&\hfil\texttt{\#Edge diff}&\hfil\texttt{EOD diff}&\hfil\texttt{Acc diff} &\hfil\texttt{F1 diff}&\hfil\texttt{\#Edge diff}&\hfil\texttt{EOD diff}&\hfil\texttt{Acc diff} &\hfil\texttt{F1 diff} &\hfil\texttt{\#Edge diff}&\hfil\texttt{Sens}&\hfil\texttt{EOD diff}&\hfil\texttt{Acc diff} &\hfil\texttt{F1 diff}    \\
\hline

\multirow{6}{*}{LR}&\multirow{2}{*}{Adult}& \multirow{2}{*}{1}  & \textbf{0.23} & -0.02 & -0.06 &  \multirow{2}{*}{0} & \textbf{0.1} & -0.01 & -0.03 & \multirow{2}{*}{3} & \textbf{0.29} & -0.09 & -0.12  &\multirow{2}{*}{2}& sex & \textbf{-0.01} & -0.01 & -0.22 \\
&&  & -0.08 & 0.0 & -0.0 &   & -0.01 & -0.0 & -0.01 &  & -0.44 & -0.03 & -0.5 &  & sex& -0.47 & -0.09 & -0.65\\\cline{2-19}

&\multirow{2}{*}{Compas} & \multirow{2}{*}{3}  & \textbf{0.06} & 0.0 & -0.0 &  \multirow{2}{*}{0} & \textbf{0.05} & -0.01 & -0.0 & \multirow{2}{*}{3} & \textbf{0.08} & -0.02 & -0.01  &\multirow{2}{*}{2}& race & \textbf{-0.02} & -0.02 & -0.01 \\
&&  & -0.04 & 0.01 & 0.0 &   & -0.03 & -0.0 & -0.0 &  & -0.07 & -0.01 & 0.0 &  & race& -0.12 & -0.02 & 0.03\\\cline{2-19}

&\multirow{2}{*}{Bank}& \multirow{2}{*}{0}  &\textbf{0.29} & -0.02 & -0.02 &  \multirow{2}{*}{0} & \textbf{0.06} & -0.0 & -0.0 & \multirow{2}{*}{1} & \textbf{0.96} & -0.03 & -0.13  &\multirow{2}{*}{2}& age & \textbf{0.0} & 0.0 & -0.37 \\
&&  & -0.01 & -0.02 & -0.01 &   & -0.01 & 0.0 & 0.01 &  & -0.06 & -0.05 & -0.1 &  &age& -0.36 & -0.44 & -0.84\\\cline{2-19}

&\multirow{2}{*}{Law School}& \multirow{2}{*}{0}& \textbf{0.05} & 0.0 & -0.0 &  \multirow{2}{*}{0} & 0.0 & 0.0 & 0.0 & \multirow{2}{*}{2} & \textbf{0.06} & -0.0 & -0.0  &\multirow{2}{*}{1}& sex & \textbf{-0.04} & -0.04 & -0.05 \\
&&  & -0.03 & 0.0 & 0.0 &   & 0.0 & 0.0 & 0.0 &  & -0.02 & -0.01 & -0.01 &  &sex & -0.1 & -0.18 & -0.07\\\cline{2-19}

&\multirow{2}{*}{Student}&  \multirow{2}{*}{0} & 0.01 & -0.0 & -0.0 &  \multirow{2}{*}{0} & \textbf{0.03} & -0.0 & -0.0 & \multirow{2}{*}{8} & \textbf{0.02} & -0.03 & -0.02  &\multirow{2}{*}{8}& sex & \textbf{-0.02} & -0.02 & -0.07 \\
&&  & -0.02 & -0.0 & 0.0 &   & -0.02 & 0.0 & 0.0 &  & -0.03 & 0.0 & 0.0 &  &sex & -0.05 & -0.09 & -0.04\\\cline{2-19}

&\multirow{2}{*}{Heart}&  \multirow{2}{*}{2} & \textbf{0.1} & 0.01 & 0.01 &  \multirow{2}{*}{0} & \textbf{0.1} & -0.0 & -0.01 & \multirow{2}{*}{2} & \textbf{0.06} & -0.03 & -0.15  &\multirow{2}{*}{1}& sex & \textbf{-0.18} & -0.18 & -0.64 \\
&&  & -0.14 & -0.01 & -0.06 &   & -0.14 & -0.01 & -0.12 &  & -0.33 & -0.04 & -0.39 &  &sex & -0.55 & -0.15 & -0.78\\\cline{2-19}
\hline

\multirow{6}{*}{DT}&\multirow{2}{*}{Adult}&  2 & \textbf{0.02} & 0.02 & 0.0 &  0 & 0.0 & -0.0 & 0.0 & 1 & \textbf{0.15} & 0.15 & 0.08  &5& sex & \textbf{0.06} & 0.06 & -0.0 \\
&&  & -0.06 & 0.06 & -0.0 &   & -0.01 & 0.01 & 0.0 &  & -0.04 & 0.04 & -0.02 &  &sex & -0.06 & 0.06 & -0.01\\\cline{2-19}

&\multirow{2}{*}{Compas}&3 & \textbf{0.06} & -0.0 & 0.15 &  0 & \textbf{0.06} & -0.01 & 0.19 & 1 & \textbf{0.06} & 0.01 & 0.24  &3& race & \textbf{0.09} & 0.02 & 0.04 \\
&&  & -0.02 & -0.01 & 0.1 &   & -0.01 & -0.01 & 0.1 &  & -0.03 & -0.03 & 0.14 &  &race & 0.01 & -0.01 & 0.03\\\cline{2-19}

&\multirow{2}{*}{Bank}& 2 & \textbf{0.05} & 0.02 & 0.0 &  0 & \textbf{0.0} & 0.0 & 0.0 & 2 & \textbf{0.11} & 0.1 & -0.11  &1& age & \textbf{0.07} & 0.07 & -0.05 \\
&&  & -0.01 & 0.01 & -0.0 &   & -0.0 & 0.0 & -0.0 &  & -0.04 & 0.04 & -0.06 &  &age & -0.03 & 0.03 & -0.05 \\\cline{2-19}

&\multirow{2}{*}{Law School}&1 & 0.02 & 0.01 & 0.0 &  0 & 0.0 & -0.0 & -0.0 & 1 & \textbf{0.03} & 0.01 & 0.08  &1& sex & 0.02 & 0.0 & 0.0 \\
&&  & 0.01 & 0.0 & -0.0 &   & -0.0 & 0.0 & -0.0 &  & -0.04 & 0.01 & 0.07 &  &sex & 0.01 & 0.0 & 0.0 \\\cline{2-19}

&\multirow{2}{*}{Student}&8 & 0.01 & -0.0 & 0.01 &  0 & 0.01 & -0.0 & 0.01 & 8 & 0.01 & -0.0 & -0.0  &8& sex & 0.0 & -0.0 & -0.0 \\
&&  & -0.0 & -0.0 & -0.0 &   & 0.0 & -0.0 & -0.0 &  & 0.0 & 0.0 & -0.02 &  &sex & -0.0 & 0.0 & -0.0 \\\cline{2-19}

&\multirow{2}{*}{Heart}& 1 & 0.0 & 0.0 & -0.0 &  0 & -0.0 & 0.0 & -0.0 & 2 & 0.04 & 0.01 & 0.02  &1& sex & 0.01 & -0.01 & 0.0 \\
&&  & -0.01 & -0.0 & 0.01 &   & -0.02 & -0.01 & 0.01 &  & 0.0 & -0.01 & 0.03 &  &sex & -0.01 & 0.01 & 0.0 \\\cline{2-19}
\hline

\multirow{6}{*}{SVM}&\multirow{2}{*}{Adult}& 1 & 0.01 & 0.01 & -0.0 &  0 & 0.0 & 0.0 & 0.0 & 5 & \textbf{0.14} & 0.13 & -0.05  &1& sex & \textbf{0.02} & 0.02 & -0.0 \\
&&  & -0.03 & -0.03 & 0.01 &   & -0.01 & -0.01 & 0.0 &  & -0.04 & 0.04 & -0.03 &  &sex & -0.03 & 0.03 & -0.0 \\\cline{2-19}

&\multirow{2}{*}{Compas}& 2 & 0.03 & 0.01 & 0.0 &  0 & 0.03 & 0.01 & 0.0 & 3 & \textbf{0.03} & 0.01 & -0.02  &2& race & 0.03 & 0.01 & 0.0 \\
&&  & -0.01 & -0.01 & -0.0 &   & -0.01 & -0.01 & 0.0 &  & -0.04 & -0.01 & -0.01 &  &race & -0.01 & -0.01 & 0.0 \\\cline{2-19}

&\multirow{2}{*}{Bank}& 1 & 0.03 & 0.03 & -0.03 &  0 & 0.01 & 0.0 & -0.0 & 2 & \textbf{0.11} & 0.1 & -0.13  &1& age & \textbf{0.08} & 0.08 & -0.04 \\
&&  & -0.01 & 0.01 & -0.01 &   & 0.0 & 0.0 & 0.0 &  & -0.07 & 0.04 & -0.08 &  &age & -0.02 & 0.01 & -0.03 \\\cline{2-19}

&\multirow{2}{*}{Law School}&  1 & 0.01 & -0.0 & -0.0 &  0 & 0.0 & 0.0 & 0.0 & 1 & 0.0 & -0.0 & -0.0  &1& sex & 0.0 & -0.0 & 0.0 \\
&&  & -0.0 & 0.0 & -0.0 &   & 0.0 & 0.0 & 0.0 &  & -0.0 & 0.0 & -0.0 &  &sex & -0.0 & -0.0 & -0.0 \\\cline{2-19}

&\multirow{2}{*}{Student}&  8 & 0.0 & 0.0 & -0.0 &  0 & 0.0 & -0.0 & 0.0 & 8 & -0.01 & -0.0 & -0.0  &8& sex & 0.0 & -0.0 & 0.0 \\
&&  & 0.0 & -0.0 & 0.0 &   & 0.0 & 0.0 & -0.0 &  & -0.01 & 0.01 & -0.02 &  &sex & -0.0 & -0.0 & 0.0 \\\cline{2-19}
&\multirow{2}{*}{Heart}& 1 & -0.0 & 0.0 & 0.0 &  0 & 0.0 & 0.0 & 0.0 & 1 & \textbf{-0.04} & -0.01 & 0.01  &2& sex & 0.0 & 0.0 & 0.0 \\
&&  & -0.01 & -0.0 & 0.0 &   & -0.01 & -0.0 & 0.0 &  & -0.1 & -0.06 & 0.04 &  &sex & -0.0 & -0.0 & 0.0 \\\cline{2-19}
\hline
\end{tabular}
}
\label{tab:table_sens_nonsens}
\end{table*}

{\footnotesize
\begin{table*}[!t]
\caption{Feature Selection and Their Fairness Characteristics over the datasets (Ablation of Causal Graphs).}
\centering
\resizebox{0.9\textwidth}{!}{%
\begin{tabular}{|c|c|c|c|c|c|c|c|c|c|c|c|c|c|}

\hline
\multirow{2}{2.5em}{\textbf{Dataset}}  &\multicolumn{3}{| c |}{\hfil\textsc{SelectKBest~\cite{KBest}}} & \multicolumn{3}{| c |}{\hfil\textsc{SelectFpr~\cite{SelectFpr}}}  &\multicolumn{3}{| c |} {\hfil\textsc{SelectPercentile~\cite{SelectPercentile}}} & \multicolumn{3}{| c |}{\hfil\textsc{DropSensParam}} \\\cline{2-13}
&\hfil\texttt{EOD diff}&\hfil\texttt{Acc diff} &\hfil\texttt{F1 diff}&\hfil\texttt{EOD diff}&\hfil\texttt{Acc diff} &\hfil\texttt{F1 diff}&\hfil\texttt{EOD diff}&\hfil\texttt{Acc diff} &\hfil\texttt{F1 diff} &\hfil\texttt{EOD diff}&\hfil\texttt{Acc diff} &\hfil\texttt{F1 diff}    \\
\hline

Adult & 0.0 & 0.0 & 0.0 & 0.0 & 0.0 & 0.0 & -0.02 & -0.02 & 0.0 & -0.09 & -0.09 & 0.0 \\ \cline{1-13}

Compas &  0.05 & 0.03 & 0.0 & 0.04 & 0.01 & 0.0 & 0.04 & 0.01 & 0.0 & -0.05 & 0.02 & 0.0 \\ \cline{1-13}
Bank &   0.03 & 0.03 & 0.0 & 0.0 & 0.0 & 0.0 & 0.03 & 0.03 & 0.01 & -0.03 & 0.03 & 0.0 \\\cline{1-13}
Law School &  0.02 & 0.02 & 0.01 & 0.02 & 0.02 & 0.01 & 0.02 & 0.02 & 0.01 & -0.03 & 0.02 & 0.0  \\\cline{1-13}
Student &  -0.02 & 0.02 & 0.0 & 0.0 & 0.0 & 0.0 & 0.0 & 0.0 & -0.01 & -0.02 & -0.02 & 0.0   \\\cline{1-13}
Heart &  0.04 & 0.04 & 0.0 & 0.0 & 0.0 & 0.0 & 0.04 & 0.04 & 0.0 & 0.05 & -0.05 & 0.0 \\\cline{1-13}
\hline
\end{tabular}
}
\label{tab:table_sens_nonsens_ablation}
\end{table*}
}

% \subsection{The systematic relationships between  sensitive \&  non-sensitive feature selection and fairness}
\vspace{0.25 em}
\noindent \textbf{Fairness and Robustness of Feature Selection via Causality (RQ2).}
\label{sec:experiments:RQ2}
% In this experiment, we aim to address the question: \emph{whether unawareness of the sensitive attribute or applying the standard (important) feature selection always results in unfair models?} 
We consider the practices of dropping a sensitive attribute shown with \texttt{DropSensParam} and
dropping non-sensitive features with \texttt{SelectKBest}~\cite{KBest}, \texttt{SelectFpr}~\cite{SelectFpr}, and \texttt{SelectPercentile} \cite{SelectPercentile}. We adjust the number of top features (k) for \texttt{SelectKBest} to exclude at most half of the features, and we use the default values of alpha=5\% and percentile=10 for \texttt{SelectFpr} and \texttt{SelectPercentile}, respectively.
The results are detailed in Table~\ref{tab:table_sens_nonsens}.
The \texttt{\#Edge diff} column shows the number of different edges between two equivalence graphs. An edge difference of $0$ implies that the same DAG graphs have distinct weights. We further assess the graphs' differences in EOD, accuracy, and F1 scores. 
% Specifically, our EOD difference measurement spans two extremes: cases where applying a certain technique significantly increases the EOD metric, and cases where it notably decreases it. 
% In our methodology, we explore the effects of dropping non-sensitive features using \texttt{SelectKBest}\cite{KBest}, \texttt{SelectFpr}\cite{SelectFpr}, and \texttt{SelectPercentile}~\cite{SelectPercentile}, alongside the practice of dropping sensitive attributes, denoted as \texttt{DropSensParam}. For \texttt{SelectKBest}, we limit the number of top features (k) to at most half of the total features. Additionally, we set $alpha=0.01$ for \texttt{SelectFpr} and $percentile=10$ for \texttt{SelectPercentile}.

\vspace{0.25 em}
\noindent \textit{Results for Dropping Sensitive Attribute.} Our findings shown in Table~\ref{tab:table_sens_nonsens} (\textsc{DropSensParam} column) highlight that the impact on fairness from dropping a sensitive attribute varies significantly, depending on the underlying causal relationships among features in a dataset. For example, in the case of the Adult dataset, dropping the gender feature leads to different outcomes in EOD. Specifically, we note a decrease of $0.13$ in the EOD for one causal graph, while a different equivalence class exhibited an increase in the EOD by $0.09$ (all within $0.01$ difference in F1 scores).
% Figure~\ref{fig:causal-bank} (a) and Figure~\ref{fig:causal-heart-drop} show two examples of neighbor causal graphs with different fairness outcomes when the sensitive attribute is dropped for Bank and Heart datasets. 

\vspace{0.25 em}
\noindent \textit{Results for selecting important non-sensitive attributes.}
The results also suggest that the causal relationships between features impact the model fairness in terms of feature selection. For example, in the Bank dataset employing \texttt{SelectPercentile} technique for excluding a set of features, our search algorithms identified two equivalence graphs with only $1$ different edge direction, one of which led to an increase in EOD by $0.15$ while the other one led to a reduction in EOD by $0.36$ (a diff of $0.21$). 
% While these equivalence graphs reduced the accuracy by at most $8\%$, we see a significant decrease in F1 by $60\%$. 
Furthermore, the impact of the underlying causal structure appeared to vary with different feature selection methods. Specifically, for the Bank dataset, the use of \texttt{SelectKBest} resulted in a $10\%$ increase in EOD. Conversely, applying \texttt{SelectFpr} on the same dataset led to a smaller EOD increase of $6\%$.  Overall, the results suggest that the property that connects selecting non-sensitive features to unfairness might not be consistently robust across different contexts. But some operators like \texttt{SelectFpr} remain robust for more benchmarks. 
Figure~9 (b-d), Figure~10, and Figure~12 show two graphs with varying fairness when applying some feature selection practice (see Appendix).

\vspace{0.25 em}
\noindent \textit{Results of analysis over the training dataset (ablating causal graphs)}.
To better understand the advantages of utilizing causal graphs, we repeat our experiments directly over the training datasets. Table~\ref{tab:table_sens_nonsens_ablation} follows a similar structure to Table~\ref{tab:table_sens_nonsens} without incorporating causal graphs.  
% where we repeat the experiments from   In Table~\ref{tab:table_sens_nonsens}, causal graphs were used to generate two neighbor datasets, allowing for the observation of significant differences in EOD across these versions. However,
When causal graphs are not used, the results demonstrate much smaller variations in EOD, accuracy, and F1 score across datasets. This limited variability suggests that, without the insights provided by causal relationships, fairness best practices appear more robust than they might be under many normative interpretations, noisy observations, faulty labeling, etc. Therefore, Table~\ref{tab:table_sens_nonsens_ablation} alone does not provide sufficient evidence to conclude the robustness of these practices.

% {\footnotesize
\begin{table*}[!t]
\caption{Impacts of Sensitive \& Non-Sensitive Feature Selection on Fairness under Distribution Shifts.}
\centering
\resizebox{\textwidth}{!}{%
\begin{tabular}{|c|c|c|c|c|c|c|c|c|c|c|c|c|c|c|c|c|c|c|}

\hline
\multirow{2}{*}{\textbf{Model}}&\multirow{2}{2.5em}{\textbf{Dataset}}  &\multicolumn{4}{| c |}{\hfil\textsc{SelectKBest~\cite{KBest}}} & \multicolumn{4}{| c |}{\hfil\textsc{SelectFpr~\cite{SelectFpr}}}  &\multicolumn{4}{| c |} {\hfil\textsc{SelectPercentile~\cite{SelectPercentile}}} & \multicolumn{4}{| c |}{\hfil\textsc{DropSensParam}} \\\cline{ 3-18}
&&\hfil\texttt{\#Edge diff}&\hfil\texttt{EOD diff}&\hfil\texttt{Acc diff} &\hfil\texttt{F1 diff}&\hfil\texttt{\#Edge diff}&\hfil\texttt{EOD diff}&\hfil\texttt{Acc diff} &\hfil\texttt{F1 diff}&\hfil\texttt{\#Edge diff}&\hfil\texttt{EOD diff}&\hfil\texttt{Acc diff} &\hfil\texttt{F1 diff} &\hfil\texttt{\#Edge diff}&\hfil\texttt{EOD diff}&\hfil\texttt{Acc diff} &\hfil\texttt{F1 diff}    \\
\hline

\multirow{6}{*}{LR}&\multirow{2}{*}{Adult}& \multirow{2}{*}{1} & \textbf{0.06} & 0.0 & 0.0 & \multirow{2}{*}{0} & 0.02 & 0.0 & 0.0 & \multirow{2}{*}{1} & \textbf{0.4} & -0.06 & -0.11 & \multirow{2}{*}{4} & \textbf{0.24} & -0.24 & -0.02\\
&&  & -0.27 & -0.02 & -0.03 &   & -0.01 & -0.0 & 0.0 &  & -0.33 & -0.13 & -0.08 &  & -0.26 & -0.02 & -0.02 \\\cline{2-18}

&\multirow{2}{*}{Compas} & \multirow{2}{*}{3} & \textbf{0.08} & -0.0 & -0.0 &  \multirow{2}{*}{0} & \textbf{0.08} & -0.0 & -0.0 & \multirow{2}{*}{2} & \textbf{0.06} & -0.02 & -0.01 & \multirow{2}{*}{3} & \textbf{0.09} & 0.05 & 0.0\\
&&  & -0.01 & -0.0 & 0.0 &   & -0.01 & -0.0 & 0.0 &  & -0.04 & -0.02 & -0.0 &  & -0.02 & 0.0 & 0.0 \\\cline{2-18}

&\multirow{2}{*}{Bank}& \multirow{2}{*}{1} & \textbf{0.04} & -0.04 & -0.04 &  \multirow{2}{*}{0} & 0.01 & 0.0 & 0.0 & \multirow{2}{*}{1} & \textbf{0.41} & -0.01 & -0.0 & \multirow{2}{*}{1} & \textbf{0.12} & 0.12 & -0.12\\
&&  & -0.12 & -0.02 & -0.07 &   & -0.01 & 0.0 & 0.0 &  & -0.28 & -0.08 & -0.37 &  & -0.06 & -0.01 & -0.02 \\\cline{2-18}

&\multirow{2}{*}{Law School}&  \multirow{2}{*}{2} & 0.04 & -0.01 & 0.0 &  \multirow{2}{*}{0} & 0.0 & 0.0 & 0.0 & \multirow{2}{*}{2} & 0.02 & -0.0 & 0.0 & \multirow{2}{*}{2} & 0.03 & 0.01 & 0.0\\
&&  & 0.01 & -0.0 & 0.0 &   & -0.0 & 0.0 & 0.0 &  & -0.02 & -0.01 & -0.01 &  & 0.01 & 0.0 & 0.0 \\\cline{2-18}

&\multirow{2}{*}{Student}&  \multirow{2}{*}{0} & 0.01 & -0.0 & 0.0 &  \multirow{2}{*}{0} & 0.01 & -0.0 & 0.0 & \multirow{2}{*}{8} & 0.01 & -0.01 & -0.0 & \multirow{2}{*}{8} & 0.01 & 0.0 & 0.0\\
&&  & -0.01 & -0.01 & -0.0 &   & -0.01 & -0.01 & -0.0 &  & -0.01 & -0.02 & -0.01 &  & -0.01 & -0.0 & 0.0 \\\cline{2-18}

&\multirow{2}{*}{Heart}&  \multirow{2}{*}{0} & \textbf{0.16} & -0.0 & 0.0 &  \multirow{2}{*}{0} & \textbf{0.12} & 0.0 & 0.0 & \multirow{2}{*}{0} & \textbf{0.26} & -0.07 & -0.06 & \multirow{2}{*}{1} & \textbf{0.16} & -0.15 & 0.02\\
&&  & -0.06 & -0.05 & -0.04 &   & -0.01 & -0.03 & -0.06 &  & 0.01 & -0.1 & -0.08 &  & -0.02 & 0.0 & 0.0 \\
\hline
\hline
\end{tabular}
\vspace{-1.0 em}
}
\label{tab:distribution_shift_LR}
\end{table*}

\vspace{0.25 em}
\noindent \textit{Robustness under distribution shifts.}
We now evaluate the robustness of feature selection techniques under conditions of distribution shift, specifically involving prior probability shifts~\cite{kull2014patterns,varshney2021trustworthy}. This shift is simulated by introducing bias terms to the label variable. We utilize different versions of the same datasets to mirror the reality of simulated concept drift. For example, the Adult dataset exhibits a 0.09 probability shift from 2015 to 2016. Consequently, we add a constant random variable, following a uniform distribution [0,$\epsilon$], to the label variable (L) to account for this shift, i.e., L $\sim$ Sigmoid($x$) $\to$ Sigmoid($x$) + $U\{0,\epsilon\}$.  
% So far, we consider in-distribution samples and carefully generate datasets that represent
% the same distribution as the input dataset. Now, we consider the robustness of some practices
% like feature selections under distribution shift. We consider only one type of distribution
% shift with prior probability shift~\cite{kull2014patterns,varshney2021trustworthy} where we
% add bias terms to the label variable (e.g., the feature $y$ in Figure~\ref{fig:Adult_DAGs} (b,c)).  
The results of this experiment for Logistic Regression are detailed in Table~\ref{tab:distribution_shift_LR} (see Table~\ref{tab:distribution_shift} in the appendix for complete results). 
Our results reveal a notable lack of robustness across all four feature selection methods when encountering a prior probability shift. 
For instance, in the Heart dataset, applying \texttt{SelectKBest}~\cite{KBest} and \texttt{SelectFpr}~\cite{SelectFpr} techniques resulted in EOD increases of $0.16$ and $0.12$, respectively. 
% This change contrasts with the previously observed stability of these methods in in-distribution samples of the same dataset (refer to Table~\ref{tab:table_sens_nonsens}). 
% A critical observation from our analysis is the identification of similar causal graphs exhibiting substantial variations in EOD, up to $0.69$, under distribution shift conditions.

\begin{answerbox}
\textbf{Answer RQ2:}
We find that removing sensitive attributes does not always degrade fairness. Also, we find that some methods of selecting non-sensitive attributes (e.g., \texttt{SelectFpr}) are more robust on model fairness than others (e.g., \texttt{SelectPercentile}). Finally, we find that the robustness of these practices varies significantly under the prior probability shifts. 
\end{answerbox}

\begin{figure}[!tb]
 \begin{minipage}{0.24\textwidth}
  \centering
  \includegraphics[width=0.795\textwidth]{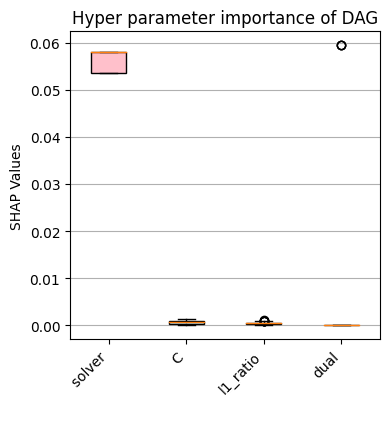}
    \captionsetup{font=small}
    \caption{HP of causal graph~\ref{fig:Adult_DAGs} (b).}
    \label{fig:base_shap_results}
 \end{minipage}%
 % \hspace{0.25 em}
 \begin{minipage}{0.24\textwidth}
  \centering
  % \vspace{1.2 em}
  \includegraphics[width=0.82\textwidth]{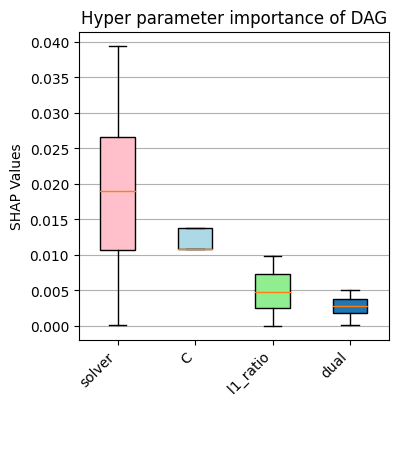}
    \vspace{-1.0 em}
    \captionsetup{font=small}
    \caption{HP of causal graph~\ref{fig:Adult_DAGs} (c).}
    \label{fig:perturbed_shap_results}
 \end{minipage}
 \vspace{-2.0 em}
\end{figure}

\vspace{0.25 em}
\noindent \textbf{Local robustness of hyperparameters for a fair design of training process (RQ3).}
\label{sec:experiments:RQ3}
We conduct a series of experiments to understand if some hyperparameters (HPs) can systematically influence
fairness. Table~\ref{tab:hyperparameters} presents the results of these experiments. 
The EOD results are averages of $30$ repeated experiments. 
Each experiment includes a $4$ hours run of an AutoML tool for fairness~\cite{10.1145/3510003.3510202},
that explores the HP space of logistic regression along with selecting and clustering
$500$ HP configurations (samples).

The column \texttt{\#Edge diff} in Table~\ref{tab:hyperparameters} indicates the number of differing edges between two equivalence causal graphs (labeled as 0 and 1). The columns HP (0) and HP (1) list the four most influential hyperparameters for two equivalent causal graphs, as identified by Shapley Additive Explanations (SHAP) analysis~\cite{NIPS2017_7062}.  In the COMPAS dataset, for example, the important HPs for graph 0 include fit\_intercept, tol, penalty, and C. In contrast, for Graph 1, the top HPs shift to tol, dual, intercept\_scaling, and max\_iteration. Notably, the hyperparameter dual is not among the top four important hyperparameters in graph 0, while it becomes the second important hyperparameter in the equivalence graph 1, indicating how the inherent causal relationships between features can alter the significance of HPs in terms of model fairness. Interestingly, some HPs like fit\_intercept consistently rank as top HPs in five out of six cases for HP (0). However, they are still not robust to similar equivalence causal graphs; in 1 out of 6, fit\_intercept is deemed significant in HP (1).
The SHAP outcome for the graph~\ref{fig:Adult_DAGs} (b) is shown in Figure~\ref{fig:base_shap_results} which illustrates the importance of four HPs where the HP `solver' has a significant impact on fairness. 
These findings may indicate that a specific set of hyperparameters is crucial in developing fair ML models.
However, when we apply SHAP on the equivalence graph~\ref{fig:Adult_DAGs} (c), we have a different set of
important hyperparameters where the same HPs like `solver' are not important. 
Thus, causal relationships between input variables are important to derive hyperparameter configurations for logistic regression, and no HPs influence fairness systematically.

% We also note that some HPs like fit\_intercept are among the top HPs in five out of six case studies in HP (0),
% but they are still not robust to similar equivalence causal graphs, one out of six cases considers fit\_intercept
% relevant in HP (1). 
% {\footnotesize
\begin{table}[b]
\vspace{-1.0 em}
% \begin{wraptable}{r}{5.40cm}
\caption{Results of hyperparameter analysis}
\label{tab:hyperparameters}
\centering
\resizebox{0.48\textwidth}{!}{
\begin{tabular}{|c|c|c|c|}
\hline
Dataset & \#Edge diff  & HP (0) & HP (1) \\
\cline{1-3}
\hline

\multirow{2}{*}{Adult} &\multirow{2}{*}{1} & solver, C,  & tol, fit\_intercept, \\
 &  & l1\_ratio, dual & intercept\_scaling, max\_iteration \\
 
\hline
\multirow{2}{*}{Compas} &\multirow{2}{*}{2} & fit\_intercept, tol, & tol, dual, \\
 &  & penalty, C & intercept\_scaling, max\_iteration \\
\hline

\multirow{2}{*}{Bank} &\multirow{2}{*}{1} & fit\_intercept, tol, & penalty, tol, \\
 &  & dual, solver & intercept\_scaling, l1\_ratio \\
\hline

\multirow{2}{*}{Law School} &\multirow{2}{*}{1} & fit\_intercept, max\_iteration, & penalty, intercept\_scaling, \\
 &  & dual, tol & max\_iteration, l1\_ratio \\
\hline

\multirow{2}{*}{Student} &\multirow{2}{*}{0} & fit\_intercept, C, & dual, C, \\
 &  & intercept\_scaling, max\_iteration & tol, l1\_ratio \\
\hline

\multirow{2}{*}{Heart} &\multirow{2}{*}{1} & penalty, dual, & max\_iteration, tol, \\
 &  & fit\_intercept, intercept\_scaling & l1\_ratio, C \\
 \hline
 
\end{tabular}
}
% \end{wraptable}
% \vspace{-1.0 em}
\end{table}
\begin{answerbox}
\textbf{Answer RQ3:}
The results show that while some hyperparameters, like fit\_intercept, consistently ranked high in importance, they did not demonstrate robustness across all causal structures. Overall, the study found no evidence to support the idea of universally "fair" or "unfair" hyperparameter selections.
\end{answerbox}

% {\footnotesize
\begin{table}[!t]
\caption{Robustness of Bias Mitigation Practices.}
\centering
\resizebox{0.48\textwidth}{!}{%
\begin{tabular}{|c|c|c|c|c|c|c|c|c|c|c|c|c|}

\hline
\multirow{2}{*}{\textbf{Model}}&\multirow{2}{2.5em}{\textbf{Dataset}}  &\multicolumn{3}{| c |}{\hfil\textsc{Treshold Optimizer~\cite{hardt2016equality}}} & \multicolumn{3}{| c |}{\hfil\textsc{Calibrated Equalized Odds}~\cite{NIPS2017_b8b9c74a}} \\\cline{ 3-8}
&&\hfil\texttt{EOD diff}&\hfil\texttt{Acc diff} &\hfil\texttt{F1 diff}&\hfil\texttt{EOD diff}&\hfil\texttt{Acc diff} &\hfil\texttt{F1 diff} \\
\hline
\multirow{6}{*}{LR}&\multirow{2}{*}{Adult}&  \textbf{0.27} & -0.07 & -0.34  & \textbf{0.07} & -0.02 & -0.07   \\
&  &  -0.1 & -0.03 & -0.07 & -0.12 & -0.02 & -0.06  \\\cline{2-8}

&\multirow{2}{*}{Compas}&  \textbf{0.03} & -0.01 & 0.0  & \textbf{0.21} & -0.02 & 0.01   \\
&  &  -0.08 & 0.0 & -0.01 &  -0.03 & 0.0 & 0.0  \\\cline{2-8}

% &\multirow{2}{*}{Bank}&  NA & NA & NA & &&   \\
% &  & NA & NA &NA &   &  &  \\\cline{2-8}

&\multirow{2}{*}{Law School}&  \textbf{0.11} & -0.02 & -0.02 & \textbf{-0.01} & -0.01 & -0.0   \\
&  &  0.06 & -0.02 & -0.01 &   -0.11 & -0.01 & -0.01 \\\cline{2-8}

&\multirow{2}{*}{Student}& 0.02 & -0.02 & -0.01 & \textbf{0.05} & -0.01 & 0.0  \\
&  & -0.02 & -0.02 & -0.01 &   -0.04 & -0.01 & 0.0  \\\cline{2-8}

&\multirow{2}{*}{Heart}&  \textbf{0.25} & -0.02 & -0.15  & 0.02 & -0.02 & -0.04   \\
&  & 0.03 & -0.04 & -0.32 &   -0.03 & -0.02 & -0.05 \\\cline{2-8}

\hline
\multirow{6}{*}{DT}&\multirow{2}{*}{Adult}&0.01 & -0.0 & -0.0  & \textbf{0.12} & 0.01 & -0.11   \\
&  & -0.01 & -0.0 & 0.0 &   -0.23 & 0.04 & -0.09 \\\cline{2-8}

&\multirow{2}{*}{Compas} & \textbf{0.02} & -0.01 & -0.01  & \textbf{0.17} & 0.06 & 0.06  \\
&  &-0.1 & 0.0 & -0.03 &   -0.11 & 0.03 & 0.03 \\\cline{2-8}

% &\multirow{2}{*}{Bank}& NA & NA& NA & &&   \\
% &  &  NA & NA & NA &   &  &  \\\cline{2-8}

&\multirow{2}{*}{Law School}& 0.01 & 0.0 & 0.0 & \textbf{-0.02} & -0.0 & 0.0   \\
&  &  -0.02 & 0.0 & 0.0 &   -0.17 & 0.01 & 0.02 \\\cline{2-8}

&\multirow{2}{*}{Student}& 0.02 & -0.01 & -0.01  & \textbf{0.08} & -0.01 & 0.0  \\
&  & -0.02 & -0.01 & -0.01 &   -0.05 & 0.0 & 0.0 \\\cline{2-8}

&\multirow{2}{*}{Heart}&  \textbf{0.07} & -0.01 & 0.02  & \textbf{0.08} & -0.01 & -0.1  \\
&  & -0.09 & 0.02 & 0.03 &   -0.13 & -0.01 & -0.06 \\\cline{2-8}
\hline

\multirow{6}{*}{SVM}&\multirow{2}{*}{Adult}& \textbf{0.2} & -0.06 & -0.27 & \textbf{0.17} & -0.01 & -0.02 \\
&  & -0.08 & -0.03 & -0.06 &   -0.22 & 0.03 & 0.4  \\\cline{2-8}

&\multirow{2}{*}{Compas}& \textbf{0.02} & -0.01 & -0.0  & \textbf{0.16} & -0.02 & 0.01   \\
&  & -0.11 & 0.0 & 0.0 &   -0.05 & -0.02 & 0.01 \\\cline{2-8}

% &\multirow{2}{*}{Bank}& NA & NA & NA & &&   \\
% &  & NA & NA & NA &   &  &  \\\cline{2-8}

&\multirow{2}{*}{Law School}&  \textbf{0.1} & -0.03 & -0.02  & 0.0 & -0.0 & 0.0  \\
&  & 0.05 & -0.01 & -0.01 &   -0.05 & -0.02 & -0.0  \\\cline{2-8}

&\multirow{2}{*}{Student}& 0.03 & -0.03 & -0.02  & \textbf{0.09} & 0.01 & 0.0   \\
&  &  -0.02 & -0.01 & -0.01 &   -0.02 & -0.01 & -0.01 \\\cline{2-8}

&\multirow{2}{*}{Heart}& \textbf{0.15} & -0.05 & -0.17 &  \textbf{0.04} & 0.0 & 0.02  \\
&  & -0.01 & -0.0 & -0.04 &   -0.19 & 0.07 & 0.46  \\\cline{2-8}

\hline
\end{tabular}
\vspace{-1.0 em}
}
\label{tab:postprocessing}
\end{table}

\vspace{0.25 em}
\noindent \textbf{Bias Mitigation Practices (RQ4).}
\label{sec:experiments:RQ4}
We examine two well-established post-processing bias mitigation algorithms: Threshold Optimizer~\cite{hardt2016equality} and Calibrated Equalized Odds~\cite{NIPS2017_b8b9c74a}. Our primary objective in this experiment is to analyze the robustness of these bias mitigation algorithms across different datasets. Results presented in Table~\ref{tab:postprocessing} where Calibrated Equalized Odds (CEO)~\cite{NIPS2017_b8b9c74a} is robust in only 2 out 15 cases, whereas Threshold Optimizer (TO)~\cite{hardt2016equality} shows robustness in 5 out of 15 cases. 
These results highlight that existing bias mitigation methods have limited local robustness. However, the effect depends on the ML algorithm and the dataset. 
For instance, when applying the LR algorithm to the Heart dataset, the CEO preserves the local robustness, whereas the TO method shows an EOD variation of  0.03 to 0.25 across neighboring datasets. Similarly, TO shows robustness with the DT algorithm trained on the Adult dataset, but the CEO fails to satisfy the property. In addition, the Student dataset with the TO method remains robust, regardless of the underlying training algorithm. 
These observations suggest that practitioners who are required to develop a fair solution may need to test the robustness using our causal search framework.

\begin{answerbox}
\textbf{Answer RQ4}: The results show that postprocessing bias mitigation practices, Threshold Optimizer~\cite{hardt2016equality} and Calibrated Equalized Odds~\cite{NIPS2017_b8b9c74a}, are not always robust locally. The robustness of these two techniques is mutually exclusive, with each solution showing superiority in distinct benchmark cases. 
\end{answerbox}

\section{Discussion}
\label{sec:discussion}

\noindent \textbf{Generative AI.}
Generative AI methods have key limitations in our setting. For example, they do not allow for the systematic exploration of the underlying data generation process, particularly the relationships between features

\vspace{0.25 em}
\noindent \textbf{Limitations.}
Our focus in this work is to test the robustness of prevalent practices in fair ML software development. We presented a novel search algorithm to explore the space of causal graphs to validate local robustness under systematic and realistic dataset shifts. The explanations of the root causes and how causal graphs structure (e.g., relationships between sensitive, non-sensitive, and outcome variables) are beyond the scope of this work. Also, our causal models assume no unobserved confounders, but hidden variables in real scenarios can affect validity.
% To overcome this limitation, methods like Fast Causal Inference (FCI)~\cite{spirtes2013causal} can be used to account for unobserved confounders and selection bias.

% several classes of causal discovery algorithms can be employed. For instance, the  algorithm can account  Using algorithms like FCI, our approach can be extended to handle unobserved and selection bias variables. Furthermore, utilizing such algorithms can further increase the accuracy of the generated data. 
% Currently focusing on logistic regression, future work should explore the generalizability of other ML algorithms. Addressing these issues by incorporating domain expertise, using non-linear models, and considering various distribution shifts will enhance the causal approach in fair and responsible ML development. 

\begin{figure}[!tb]
 \begin{minipage}{0.24\textwidth}
  \centering
  \includegraphics[width=0.795\textwidth]{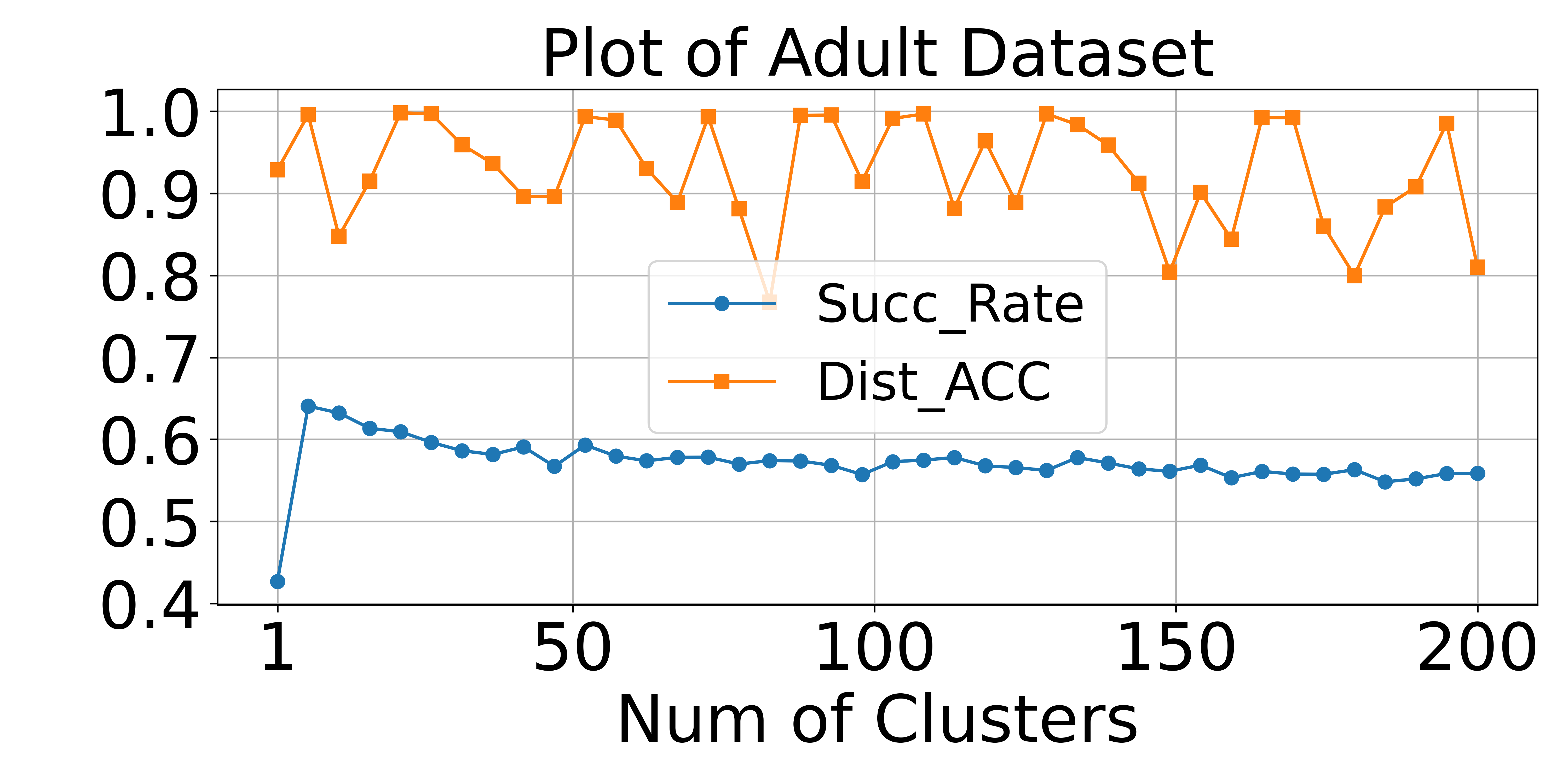}
  \includegraphics[width=0.795\textwidth]{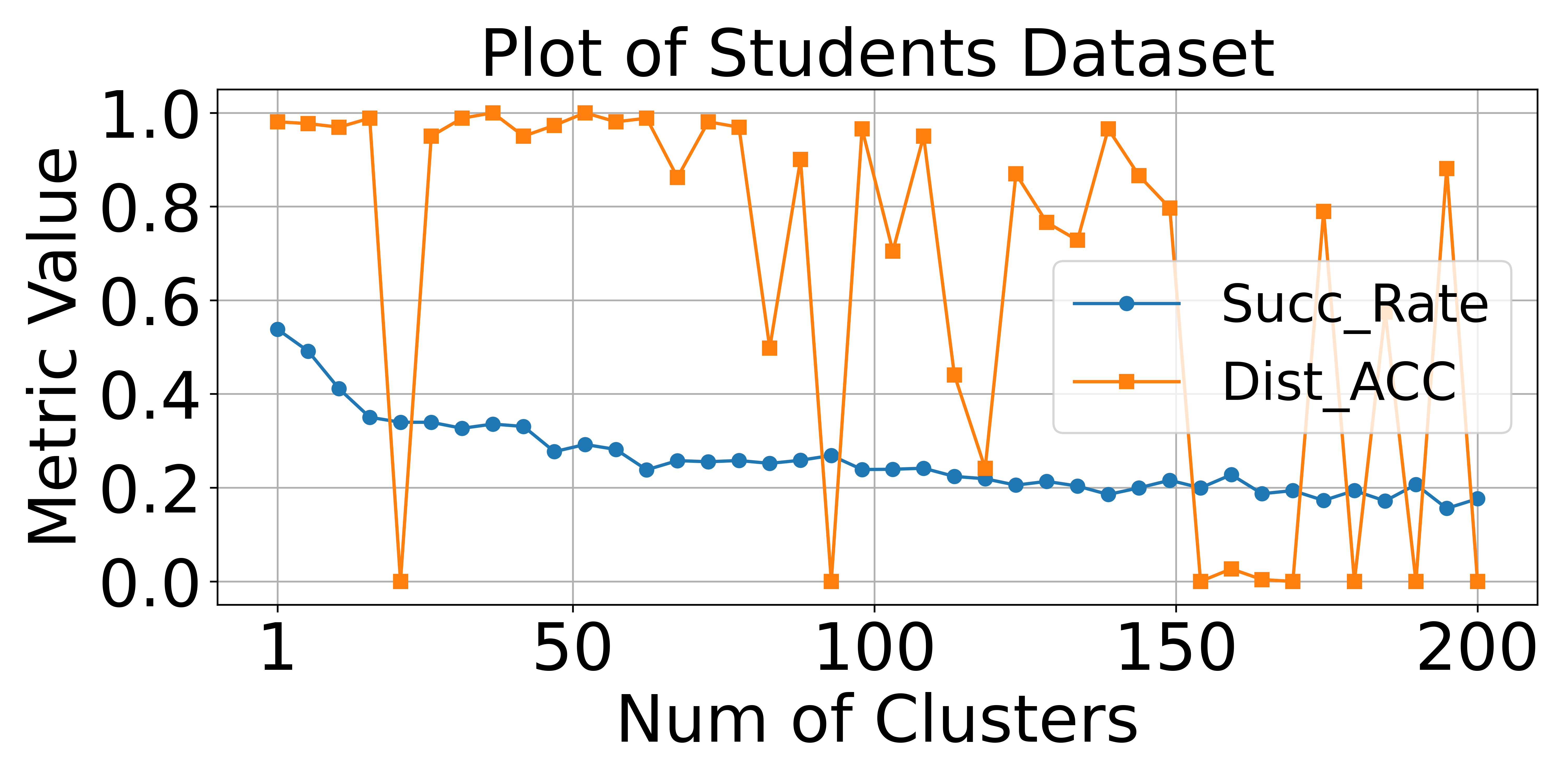}

 \end{minipage}%
 \begin{minipage}{0.24\textwidth}
  \centering
  \includegraphics[width=0.795\textwidth]{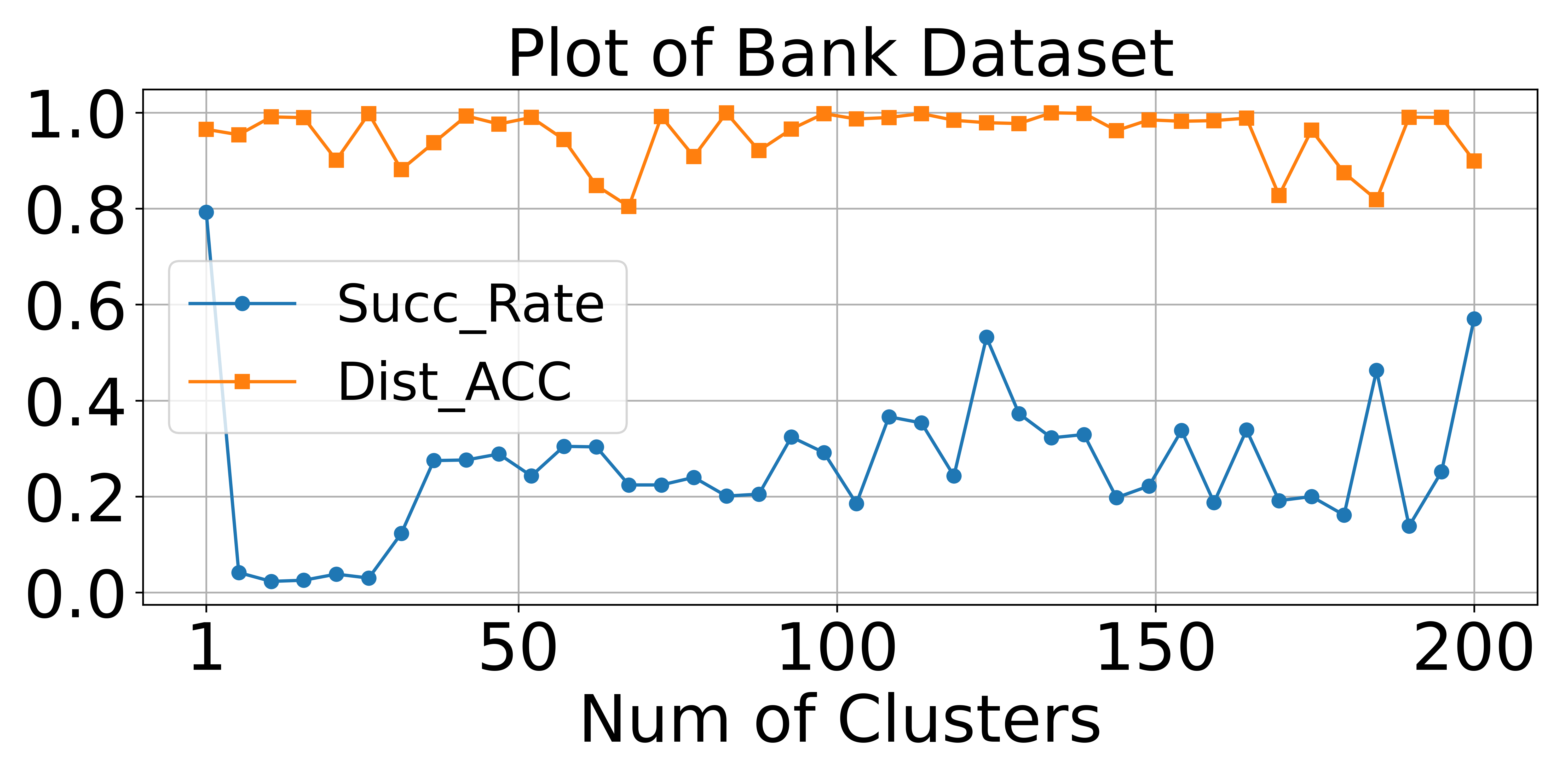}
  \includegraphics[width=0.795\textwidth]{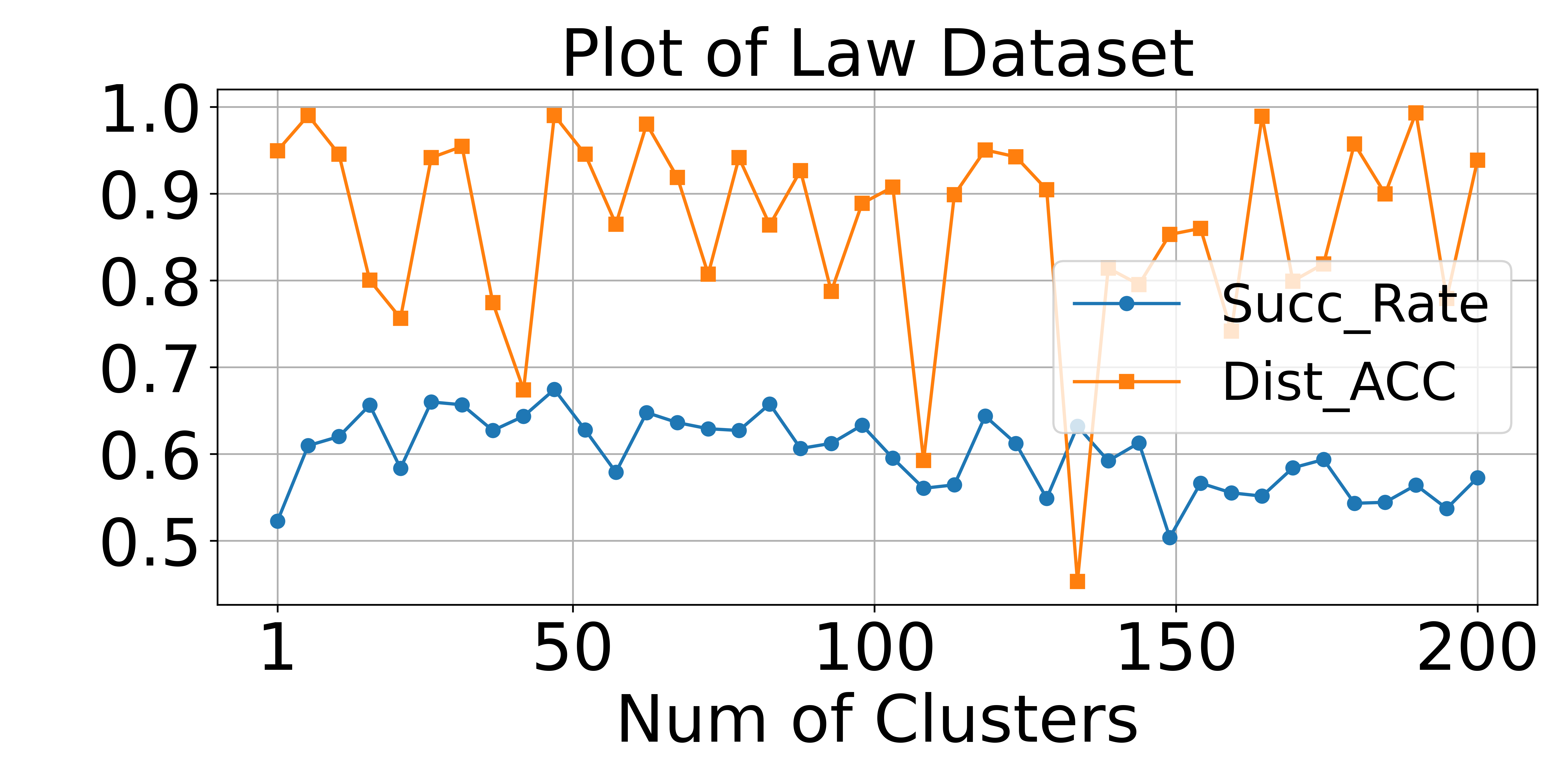}

 \end{minipage}%
\caption{Sensitivity Analysis of Cluster Numbers.}
\label{fig:cluster_analysis}
\vspace{-2.0 em}
\end{figure}

\vspace{0.25 em}
\noindent \textbf{Threat to Validity.} To address the internal validity and ensure our findings do not lead to invalid conclusions, we followed established guidelines and used the Scott-Knott statistical testing to convey significant results. To assess the impact of the number of clusters (set to 100 in our experiments for all datasets) on the performance of the distance function and the success rate of causal graphs, we varied it from 1 to 200 across datasets as shown in Figure \ref{fig:cluster_analysis}.  
While the results show that ideal values depend on the datasets and can be judiciously chosen to maximize success rates, our (global) design choice is still reasonably close to the ideal values.
% We exclude the Bank dataset~\cite{Dua:2019-bank} from the post-processing experiments, as it has insufficient privileged group samples with unfavorable labels required by the bias mitigators.
To ensure that our results are generalizable, we used six datasets, three training algorithms, three causal discovery algorithms, and eight design patterns. We utilized the EOD notion of fairness, which, while effective, might overlook unfairness detectable through fairness definitions. However, our approach is adaptable and can be applied to additional fairness metrics, such as AOD, SPD, and DI fairness.
Linear models for causal inference may not fully reflect the complexity of real-world variable relationships. 
% To scale our approach to a complex setting, one can explore other MCMC algorithms like Hamiltonian Monte-Carlo~\cite{betancourt2017conceptual}.
% We only studied logistic regression classifier, since it has been a default choice in the literature of SE for fairness. But, it might not be representative of all other classifiers. 
% We selected a subset of prevalent practices (e.g., feature important selections and hyperparameter configurations) from the state-of-the-art SE venues (ICSE, FSE, ASE, and ISSTA) in the past 5 years. But, we may miss some other important practices from other literature.
% We also applied one type of distribution shifts, guided by the prior distribution shift of Adult dataset.
% This might not reflect the shift from other datasets, and further research is needed to study other types of distribution shifts.  

\noindent
\textbf{Intended Use and Practical Workflow.} Our framework is intended for SE and ML practitioners, building fairness-critical components within data-driven software systems. 
% As fairness interventions become increasingly standardized---such as dropping sensitive features, adjusting hyperparameters, or applying post-processing mitigations---a key challenge remains: do these interventions remain effective under plausible shifts in the data-generating process? 
Our framework provides a pre-deployment testing mechanism to address the robustness of fairness interventions. Here is the \textit{workflow:}
\begin{itemize}
    \item \textbf{Input.} The practitioner provides the original training dataset and chooses a group fairness metric of interest (e.g., demographic Parity). They then specify the fairness intervention they wish to evaluate (e.g., feature selection).

    \item \textbf{Process.} Using the training data, the framework automatically constructs a causal graph representing the underlying relationships between variables. It then algorithmically searches the space of causally-equivalent graphs to identify a "neighboring" data distribution. The goal is to find a plausible data variation where a fairness intervention results in a significant degradation of the fairness metric. 

    \item \textbf{Output.} If the intervention's effectiveness degrades under these local variations, the tool flags the practice as "non-robust," alerting the practitioner that its benefits may not be reliable in production.
\end{itemize}

% \begin{figure}[!tb]
%  \begin{minipage}{0.24\textwidth}
%   \centering
%   \includegraphics[width=0.795\textwidth]{Figures/Student_cluster_analysis.png}
%     \captionsetup{font=small}
%     \caption{Num Clusters Analysis.}
%     \label{fig:student_cluster}
%  \end{minipage}%
%  \begin{minipage}{0.24\textwidth}
%   \centering
%   \includegraphics[width=0.795\textwidth]{Figures/Law_cluster_analysis.png}
%     \captionsetup{font=small}
%     \caption{Num Clusters Analysis.}
%     \label{fig:law_cluster}
%  \end{minipage}%
% \end{figure}

\noindent 
\textbf{Implications for Software Engineering.} This paper advocates for a conceptual shift in how the SE community approaches algorithmic fairness: from treating it as a static property to viewing it as a dynamic requirement that must be continuously validated.

\vspace{0.25 em}
\noindent \textit{Implications for SE Research.} By operationalizing fairness robustness as a testable property, our work opens several new research avenues at the intersection of fairness, testing, and reliability. It provides a foundation for developing novel techniques in areas such as regression testing---creating test suites that automatically check if changes to underlying data degrade fairness guarantees.

\vspace{0.25 em}
\noindent \textit{Implication for SE Practitioners. } 
Our framework empowers engineers to proactively stress-test fairness interventions before deployment, much like they already do for security and performance. Fairness, like security, must be treated as a robustness concern in SE, requiring testing under varying conditions.

\section{Related Work}
\label{sec:related}

\noindent \textbf{Empirical Recommendations on Fair Designing Training Process.}
Zhang and Harman~\cite{zhang2021ignorance} found that enlarging the feature space during training can improve fairness while increasing the size of samples does not affect fairness. \textsc{Fairway}~\cite{chakraborty2020fairway,chakraborty2019software} showed that the hyperparameter tuning
can help mitigate the bias of data-driven software. Nguyen et el.~\cite{nguyen2023fix} used AutoML techniques~\cite{weerts2023fairness} to improve fairness with minimal degradation of functional accuracy. 
Crucially, \textsc{Parfait-ML}~\cite{10.1145/3510003.3510202}
found that some hyperparameter configurations can systematically introduce fairness bugs in the data-driven software. Gohar et al.~\cite{gohar2023understanding} recently extended this to understand how ensembles of ML models and their hyperparameters influence fairness. Biswas and Rajan~\cite{10.1145/3468264.3468536} studied how different data preprocessing stages impact fairness by excluding/including one operator while keeping every other operator the same. 
% For example, they confirm that selecting a subset of features often increases unfairness, but the effects depend on the type of operator, e.g., they reported that \texttt{SelectFpr} that chose
% features based on false positive rates do not impact fairness, while \texttt{SelectBest} that pick top $k$
% most important features often increase unfairness. 
We systematically study these findings to understand their local robustness. Recently, Monjezi et al.~\cite{10.1145/3639478.3643530} advocate for using causal graph fuzzing to probe the robustness of fairness practices. 
While we share the goal of using causal graphs for fairness analysis, our approach differs fundamentally in methodology and scope. Their proposed method relies on fuzzing—randomly perturbing the causal graph—to generate variant datasets. In contrast, our framework performs a principled search across a formally defined space of causal equivalence classes. This avoids generating implausible or invalid data distributions that can arise from random perturbations. Furthermore, their empirical study is a preliminary exploration of a single dataset and intervention. We significantly advance this line of work by developing a fully automated robustness testing framework and conducting a large-scale evaluation across multiple datasets, fairness interventions, and learning algorithms.

\vspace{0.25 em}
\noindent \textbf{Causality and Fairness.} The ML community has extensively explored fairness using causality concepts~\cite{Counterfactual-Fairness,kilbertus2017avoiding,galhotra2022hyper,10.5555/3060832.3061001}. Kusner et al.~\cite{Counterfactual-Fairness}
leveraged counterfactual reasoning to augment data samples with values from unobserved
variables and then infer linear models to predict outcomes without using any protected variables
or their ascendants in the causal graph. Zhang et al.~\cite{10.5555/3060832.3061001} employed causal Bayesian networks (CBN) for situation testing to find similar inputs and measure distances based on each attribute's causal impact on outcomes. They identified dataset discrimination if two groups from different backgrounds received notably different outcomes. 
Ji et al. \cite{10.1109/ASE56229.2023.00105} use causal analysis to explore the inherent trade-offs between fairness and other critical system metrics, such as model accuracy. While they use causality to model the relationships between different metrics, we use causality to model plausible variations in the underlying data distribution. Our aim is not to analyze trade-offs, but to determine if a given fairness intervention is fragile and likely to fail when faced with realistic data shifts.

% \textsc{Fair-SMOTE}~\cite{Chakraborty-FSE'21} used situation testing to identify and correct biases in the training dataset arising from incorrect labeling. They compared two similar individuals from different communities with differing outcomes to address biases without merely altering protected attributes. 

% \textsc{Fair-SMOTE}~\cite{Chakraborty-FSE'21} used
% situation testing to detect and mitigate biases in the training dataset caused by faulty labeling. They searched for two similar individuals (in the dataset) from two different communities that receive different outcomes (rather than blindly perturbing protected attributes). The vast majority of related work incorporates causal reasoning to
% detect and mitigate biases in the dataset or model. Instead, we incorporate the causal model of data
% for a fair design of training process.

\vspace{0.25 em}
\noindent \textbf{Causality in Fairness Testing and Debugging.} The notion of individual discrimination~\cite{10.1145/3106237.3106277} has been significantly used to test software for
discrimination~\cite{angell2018themis,10.1145/3338906.3338937,agarwal2018automated,udeshi2018automated,10.1145/3510003.3510137,zhang2020white,10.1145/3460319.3464820,9793943}. 
\textsc{Themis}~\cite{angell2018themis} measures the difference in outcomes between a group of individuals with the
protected attributes $A$ and a counterfactual group of the same individuals whose protected attributes are set to $B$.
 DICE~\cite{10.1109/ICSE48619.2023.00136} uses an information-theoretic approach to identify individual-level fairness violations and localize the specific neurons or layers within a deep neural network responsible for them. Its primary focus is on debugging the internal mechanics of the model itself. Rather than testing model fairness, we test the robustness of fairness practices.

\section{Conclusion}
\label{sec:conclusion}
% Data-driven software development is a critical step
% in approaching challenging societal problems in
% domains such as autonomous driving, finance, admissions, parole
% decisions, and tax auditing. However, due to its data-driven nature, 
% ML software can inadvertently perpetuate societal biases, 
% leading to disproportionate impacts on various racial, gender, or age groups.
% To improve fairness of ML software and reduce its potential harms,
% the ML engineering community has been proactive in establishing best practices for
% the development of fair ML software, such as careful feature selection and
% specific hyperparameter configurations. However, translating and validating
% these rule-of-thumb practices have remained a challenge. 
% Our study challenged the universality of these best practices, positing that
% the robustness of these practices across different settings is crucial to their validity. 
% Over a set of six case studies,
% we found that the practices are valid for some tasks, but not for others,
% depending on the underlying causal relationships between variables.
% For future work, we plan to investigate the causal theory (interventions
% and counterfactuals) to provide intuitive explanations under which a practice
% may or may not improve fairness. 

The SE community has established best practices for fair ML development, such as careful feature selection, hyperparameter tuning, and bias mitigation, but translating and validating these rule-of-thumb practices remains challenging.
Our study reveals that the effectiveness of these practices varies across different settings depending on the underlying causal relationships between variables. 
Our causal testing framework can enable other research to assess the local robustness of their proposed algorithms for in-distribution (e.g., noisy observations) and out-of-distribution (e.g., label shifts). We plan to explore causal theory to explain when and why certain practices improve fairness for future work.

\noindent \textbf{Acknowledgment.}
This project has been partially supported by NSF under grants CNS-2527657, CNS-2230061, CCF-2532965, and CCF-2317207.

\newpage

\bibliographystyle{ACM-Reference-Format}
\bibliography{reference}

% \clearpage

% \input{Sections/Appendix}
% \input{Sections/Table_RQ2_OOD}

\end{document}